\documentclass[aps,prd,twocolumn,superscriptaddress,longbibliography]{revtex4-2}
\usepackage{graphics}
\usepackage[normalem]{ulem}
\usepackage{amssymb}
\usepackage{amsmath}
\usepackage{bbm}
\usepackage{amsfonts}
\usepackage{color}
\usepackage{graphicx}
\usepackage{amsthm}
\usepackage{paralist}
\usepackage{verbatim}
\usepackage{chemarr}
\usepackage{braket}
\usepackage[dvipsnames]{xcolor}
\allowdisplaybreaks

\def\>{\rangle}
\def\<{\langle}

\usepackage{hyperref}

\begin{document}

\title{Experimental Quantum Communication Enhancement by Superposing Trajectories}

\author{Giulia Rubino}
\thanks{Corresponding Author: giulia.rubino@univie.ac.at}
\affiliation{Vienna Center for Quantum Science and Technology (VCQ), Faculty of Physics, University of Vienna, Boltzmanngasse 5, 1090, Vienna, Austria}

\author{Lee A. Rozema}
\affiliation{Vienna Center for Quantum Science and Technology (VCQ), Faculty of Physics, University of Vienna, Boltzmanngasse 5, 1090, Vienna, Austria}

\author{Daniel Ebler}
\affiliation{Institute for Quantum Science and Engineering, Department of Physics,
Southern University of Science and Technology (SUSTech), Shenzhen, China}
\affiliation{Wolfson College, University of Oxford, Linton Road, Oxford, United Kingdom}
\affiliation{QICI Quantum Information and Computation Initiative, Department of Computer Science, The University of Hong Kong, Pok Fu Lam Road, Hong Kong}

\author{Hl\'er Kristj\'ansson}
\affiliation{Quantum Group, Department of Computer Science, University of Oxford, Wolfson Building, Parks Road, Oxford, OX1 3QD, United Kingdom}
\affiliation{HKU-Oxford Joint Laboratory for Quantum Information and Computation}

\author{Sina Salek}
\affiliation{Fujitsu Laboratories of Europe, 4th Floor, Building 3, Hyde Park Hayes, 11 Millington Road, Hayes, Middlesex, UB3 4AZ, United Kingdom}

\author{Philippe Allard Gu\'erin}
\affiliation{Vienna Center for Quantum Science and Technology (VCQ), Faculty of Physics, University of Vienna, Boltzmanngasse 5, 1090, Vienna, Austria}
\affiliation{Institute for Quantum Optics and Quantum Information (IQOQI), Austrian Academy of Sciences, Boltzmanngasse 3, 1090 Vienna, Austria}

\author{Alastair A. Abbott}
\affiliation{Department of Applied Physics, University of Geneva, 1211 Geneva, Switzerland}

\author{Cyril Branciard}
\affiliation{Univ. Grenoble Alpes, CNRS, Grenoble INP, Institut Néel, 38000 Grenoble, France}

\author{\v Caslav Brukner}
\affiliation{Vienna Center for Quantum Science and Technology (VCQ), Faculty of Physics, University of Vienna, Boltzmanngasse 5, 1090, Vienna, Austria}
\affiliation{Institute for Quantum Optics and Quantum Information (IQOQI), Austrian Academy of Sciences, Boltzmanngasse 3, 1090 Vienna, Austria}

\author{Giulio Chiribella}
\affiliation{QICI Quantum Information and Computation Initiative, Department of Computer Science, The University of Hong Kong, Pok Fu Lam Road, Hong Kong}
\affiliation{Quantum Group, Department of Computer Science, University of Oxford, Wolfson Building, Parks Road, Oxford, OX1 3QD, United Kingdom}
\affiliation{HKU-Oxford Joint Laboratory for Quantum Information and Computation}
\affiliation{Perimeter Institute for Theoretical Physics, 31 Caroline Street North, Waterloo, Ontario, Canada}

\author{Philip Walther}
\affiliation{Vienna Center for Quantum Science and Technology (VCQ), Faculty of Physics, University of Vienna, Boltzmanngasse 5, 1090, Vienna, Austria}
\affiliation{Research Platform TURIS, University of Vienna, Boltzmanngasse 5, 1090, Vienna, Austria}

\begin{abstract}
%
In quantum communication networks, wires represent well-defined trajectories along which quantum systems are transmitted.
%
In spite of this, trajectories can be used as a quantum control to govern the order of different noisy communication channels, and such a control has been shown to enable the transmission of information even when quantum communication protocols through well-defined trajectories fail.
%
This result has motivated further investigations on the role of the superposition of trajectories in enhancing communication, which revealed that the use  of quantum-control of parallel communication channels, or of channels in series with quantum-controlled operations can also lead to communication advantages.
Building upon these findings, here we experimentally and numerically compare different ways in which two trajectories through a pair of noisy channels can be superposed.
We observe that, within the framework of quantum interferometry, the use of channels in series with quantum-controlled operations generally yields the largest advantages.
Our results contribute to clarify the nature of these advantages in experimental quantum-optical scenarios, and showcase
the benefit of an extension of the quantum communication paradigm in which both the information exchanged and the trajectory of the information carriers are quantum.
\end{abstract}
  
\maketitle

\section{Introduction}
The ability to establish secure communication linkages is of paramount importance in any information technology. 
Quantum cryptography protocols \cite{Bennett1984,ekert1991} achieve this in a stunning way, enabling a sender and receiver to communicate securely even in the presence of an eavesdropper with unlimited computational power.
The crucial ingredient for this feat is the availability of reliable transmission lines for quantum particles.
In this framework, any noisy process affecting the transmission is attributed to the presence of an eavesdropper, and when the noise exceeds a given threshold, the security of the communication is considered compromised. For this reason, the mitigation of any noise arising from faulty transmission lines is an integral part of the efforts to enable secure communication.

Within the quantum communication networks paradigm, quantum communication protocols encode information in quantum states, yet they treat the propagation of information carriers as classical \cite{Chiribella2009_quantumNetworks}.
Nevertheless, the information carriers can propagate along non-classical trajectories, experiencing a coherent superposition of alternative quantum evolutions  \cite{Aharonov1990, aaberg2004subspace, Oi2003}. Taking advantage of this fact, Gisin \textit{et al.} \cite{gisin2005} realized that quantum superpositions of trajectories can be harnessed to reduce the noise induced by a pair of noisy communication channels.
Therein, it was shown that when the quantum information carriers~\footnote{We use the notion of `particles' as a synonym for quantum systems which, naturally, can be delocalised in space and time. These quantum systems are used as carriers of quantum information, and in this sense we interchangeably refer to them also as `information carriers'.} are sent through two noisy channels in a quantum superposition of trajectories, interference between the two resulting noisy processes can sometimes lead to partial cancellation of the noise via post-selection.

Recently, interest in this discovery has been revived by studies emerging from quantum foundations.
In particular, it was shown that the superposition of trajectories can generate setups where the order of different channels is in a quantum superposition. These setups produce the same output as a mathematical map called the `quantum switch' \cite{chiribella2009_preprint,Chiribella2013}, a higher-order operation which takes two quantum channels as input and combines them in a quantum-controlled order. The quantum switch is an instance of
a causally-indefinite process; such processes are currently the target of wide-ranging research both for fundamental reasons \cite{Hardy_2007, Oreshkov2012, Baumeler_2016}, and for their potential to provide advantages in quantum computation \cite{Hardy2009,Chiribella2012,COLNAGHI20122940,Chiribella2013,Facchini2015,Araujo2017,Taddei2020}, quantum communication complexity \cite{baumeler2014,feix2015,Guerin2016}, and quantum metrology \cite{zhao2019}.
Moreover, the particular class of causally-indefinite processes  based on the superposition of alternative orders  can be probed via current experimental technologies, as has been recently done by encoding information in various degrees of freedom of single photons \cite{procopio2015,rubino2017,rubino2017experimental,goswami2018,wei2019experimental,goswami2018communicating,guo2020experimental,Taddei2020}.

It was further proposed \cite{Ebler2018, salek2018quantum, chiribella2018indefinite} that the quantum switch can also reduce noise in classical and quantum communication.
These findings triggered a host of subsequent proposals~\cite{Procopio2019, procopio2020sending, caleffi2020, Sazim2020, Wilson2020}, and even a few experiments~\cite{ goswami2018communicating,guo2020experimental}, highlighting the advantage of using quantum superpositions of noisy channels in alternative orders to reduce transmission noise. 

However, alongside the body of work focused on superpositions of alternative orders, the use of superpositions of trajectories in quantum communication has also been investigated~\cite{abbott2018,chiribella2019quantum,guerin2019,kristjansson2019,Grinbaum2020}. In this context, theoretical studies have pointed out that causal-indefiniteness is not necessarily required to reduce the noise in classical and quantum communication~\cite{abbott2018,guerin2019,Grinbaum2020}.
In particular, similar or even better advantages can be achieved by using a quantum-control of parallel noisy channels~\cite{abbott2018}, or by placing channels in series with quantum-controlled operations~\cite{guerin2019}. Indeed, in Ref.~\cite{guerin2019} it was even shown that the Shor quantum error correcting code can be used to find a channel layout in series with quantum-controlled gates which allows any arbitrary noise to be completely eliminated. 
%
%
This suggested the need for a thorough information-theoretic understanding of the resources in play, and a unified description of such protocols. 
One such approach is presented in Refs.~\cite{kristjansson2019,chiribella2019quantum}.
On the other hand, the comparison of different protocols can be also viewed as an experimental task, wherein one wishes to classify and quantify the experimental resources required for a physical implementation of the various types of superpositions of trajectories and their corresponding advantages \footnote{In this paper, in contrast to Ref.~\cite{kristjansson2019}, we consider resources as an experimental concept, corresponding to the physical devices and their interactions as they occur in the laboratory.}.

\begin{figure}
  \centering%
      \includegraphics[width=.95\columnwidth]{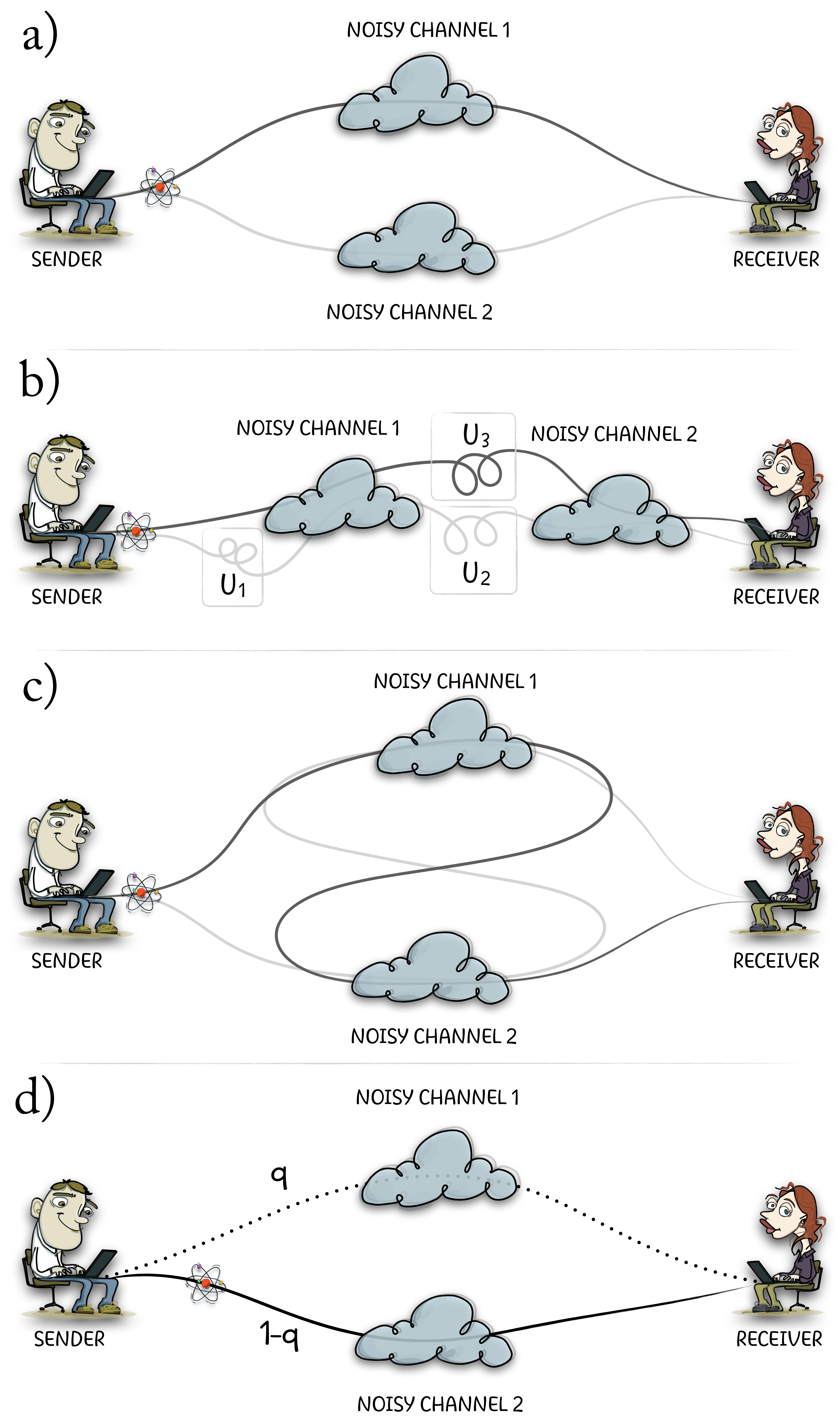}
  \caption{\textbf{Combining two channels in a superposition of trajectories.} A sender and a receiver communicate under the restriction that the information carrier must cross at least one noisy region.
	\textbf{a)} \textbf{Quantum-Control of Parallel Channels}. A quantum particle is placed in a quantum superposition of two trajectories, each branch containing a single noisy channel.
	\textbf{b)} \textbf{Channels in Series with Quantum-Controlled Operations.} Each of the branches of the superposition passes through the noisy channels in the same order, but different unitary operations are applied locally in each branch.
	\textbf{c)} \textbf{Quantum-Control of Channel Order.} The information carriers are routed through the two channels in different orders. This setup can achieve a genuinely indefinite order of the two channels.
	\textbf{d)} \textbf{Classical Trajectories.} Throughout this article, we will compare the three quantum superpositions of channels above to classical trajectories. In this regard, if one has access to classical-like trajectories only, one can send the photon through one or the another noisy regions with probabilities $q$ and $1-q$.}
    \label{cartoon}
\end{figure}

We take the experimental approach here, focusing on three different types of superpositions of trajectories which have been identified in the literature, namely, quantum-control of parallel channels (Fig.~\ref{cartoon}\textbf{a)}), channels in series with quantum-controlled operations (Fig.~\ref{cartoon}\textbf{b)}), and quantum-control of channel order (Fig.~\ref{cartoon}\textbf{c)}). While previous experimental studies~\cite{goswami2018communicating,guo2020experimental} focused only on the reduction of noise with an indefinite causal order, no experimental work had so far implemented the other proposed schemes, nor had they compared them with indefinite causal structures to provide an exhaustive assessment of the resources in play. We find that the common resource in all the three schemes considered is the establishment of a coupling between the trajectories of the information carriers and the degree of freedom on which the noise acts. On this basis, we propose a fundamentally new understanding of the resources required for this noise reduction than that proposed in previous experimental works in this field~\cite{goswami2018communicating,guo2020experimental}.

We experimentally apply the above three schemes to various noise models. 
This enables us to examine the utility and trade-offs of these different types of superpositions in the goal of communicating through a pair of noisy channels. 
In particular, in order to perform a comparative analysis of the performance of the three types of superpositions, we measure the coherent information (which is a lower bound for the quantum channel capacity) in the presence of \textit{XY}, bit-flip, phase-flip and BB84-channels.
We show that, within the paradigm of quantum interferometry, the use of channels in series with quantum-controlled operations generally outperforms or equals the other schemes in all the noise models which we consider. While here we study the three schemes individually in order to focus on the source of the coupling between the trajectory and the degree of freedom on which the noise acts, one could of course also combine the different types of superpositions (and, for instance, insert quantum-controlled operations also in the other two schemes), yielding different---potentially larger---advantages from those presented here.

The rest of this paper is structured as follows.  
Section \ref{sec:QSuperp_Trajectories} introduces the three different architectures for the quantum superpositions of trajectories through two noisy channels, and summarizes their performance when applied to a simple noise model.  
Section \ref{sec:Quantif_Channel_Performance} reviews the key figures of merit that we use to quantify the performance of our experimental quantum channels, \textit{i.e.}, the quantum capacity and the coherent information.
Section \ref{sec:experiment} outlines our experiments, and Section \ref{sec:Results} presents the corresponding results. Finally, Section \ref{sec:Discussion} concludes.

\section {Quantum Superpositions of Trajectories}
\label{sec:QSuperp_Trajectories}

For simplicity, we will focus on two communication channels and two trajectories, as this already captures the key features of the general idea. 

All experiments hereafter discussed were performed using single photons, where the trajectory is naturally defined by the photon's path.
Quantum information is initially encoded in one of the internal degrees of freedom of the particle (we refer to Section~\ref{subsec:classical} for a discussion of the case of classical information); in our case, in its polarization.
Then, using linear optical elements, it is relatively easy to place a photon in a superposition of trajectories \cite{procopio2015,rubino2017,rubino2017experimental,goswami2018}.
We will further consider, as does related work, that the noise acts only on the internal degree of freedom (DOF).

To introduce the basic idea, we will start by considering a particular noise model, which was studied for quantum-controlled orders in \cite{chiribella2018indefinite}.
Given some single-qubit input state $\rho$ encoded in the internal DOF, the noisy process $\mathcal{C}$ either applies a Pauli-$X$ or -$Y$ operation to the internal state with equal probability:
\begin{equation}\label{eq:XYsimp}
\mathcal{C}_{\text{EB}}(\rho)=\frac{1}{2} X\rho X + \frac{1}{2} Y\rho Y.
\end{equation}
If the input to this process is a pure state $\ket{\psi}_\mathrm{I}=\alpha\ket{0}_\mathrm{I}+\beta\ket{1}_\mathrm{I}$ (where the subscript I denotes the internal DOF), the output is, in general, a mixed state, with all coherence in the computational basis extinguished:
\begin{equation}
\mathcal{C}_{\text{EB}}(\ket{\psi}_\mathrm{I}\bra{\psi}_\mathrm{I})=
\begin{pmatrix}
|\beta|^2 & 0\\
0 & |\alpha|^2
\end{pmatrix},
\end{equation}
and as such, it cannot be used to transmit any quantum information.
One might, of course, still employ it to transmit classical information in the computational basis.
This channel is an example of a so-called `entanglement-breaking' (EB) channel, which would destroy any preexisting entanglement between the transmitted qubit and any other system.

In a standard quantum communication scenario with a single trajectory, information, which is taken to be encoded in an internal DOF of an information carrier, must often propagate through multiple channels.  
Depending on the physical implementation, the channels can be linked together in different manners.
With two channels and classical-like trajectories, the channels can either be put in series, or in a classical mixture of the two (depicted in Fig.~\ref{cartoon}\textbf{d)})---more complex combinations can also be realized, but they all perform strictly worse than a classical mixture.
If two copies of the channel of Eq.~\eqref{eq:XYsimp} are put in series, the result is a maximally-dephasing channel $\mathcal{C}(\rho)=\frac{1}{2}\rho + \frac{1}{2}Z \rho Z$, where $Z$ is the Pauli-$Z$ matrix. This also destroys all coherence in the computational basis, and cannot transmit any quantum information. Similarly, placing two of these channels in a classical mixture will not allow the transmission of any quantum information.

In a typical single-trajectory quantum communication scenario, it can be shown that, if each channel is unable to transmit quantum information (\textit{i.e.}, its quantum capacity---to be defined later---is zero), then any combinations of the two channels should also result in a zero capacity channel. This is known as a bottleneck inequality \cite{leditzky2018dephrasure}.
In the following subsections, we will show that this is not the case when the trajectories are superposed in a quantum fashion. 
Thus, the bottleneck inequality does not directly apply to communication scenarios with quantum trajectories \cite{salek2018quantum,guerin2019}.

\subsection{Quantum-Control of Parallel Channels}
\label{subsec:QCPC}

The first layout that we consider uses a quantum superposition of configurations where two independent channels are placed in parallel, and their use is controlled by a quantum system, as illustrated in Fig.~\ref{cartoon}\textbf{a)}.
This was originally introduced for error filtration \cite{gisin2005}, and it was more recently reviewed in the general framework of communication through superposed channels in Refs.~\cite{chiribella2019quantum, abbott2018}.
In this scheme, different independent noisy channels are placed in each branch of the superposition.
In Ref.~\cite{gisin2005}, it was shown that by performing a measurement on the trajectory in a suitable basis, and then post-selecting, one can non-deterministically filter out errors in the communication channel.
We will now consider an initial pure state encoded in the internal DOF $\ket{\psi}_\mathrm{I}=\alpha\ket{0}_\mathrm{I}+\beta\ket{1}_\mathrm{I}$, independent noisy channels realised by applying a Pauli-$X$ and -$Y$ with equal probabilities---as described previously and resulting in Eq.~\eqref{eq:XYsimp}--- and two trajectories in an equal superposition $\ket{+}_\mathrm{T} = \bigl(\ket{0}_\mathrm{T} + \ket{1}_\mathrm{T}\bigr)/\sqrt{2}$ (where T refers to the trajectory DOF).
It is then straightforward to calculate the output (the full calculation is presented in Section \ref{subsec:calc}), and to observe that performing a measurement on the trajectory DOF in the $\{ \ket{+}_\mathrm{T},\ket{-}_\mathrm{T}\}$ basis, and finding $\ket{-}_\mathrm{T} = \bigl(\ket{0}_\mathrm{T} - \ket{1}_\mathrm{T}\bigr)/\sqrt{2}$ (which,  as shown in Section \ref{subsec:calc}, occurs with probability 1/4) leaves the internal DOF in the pure state:
\begin{equation}\label{eq:minusPS}
\beta\ket{0}_\mathrm{I} -i\alpha\ket{1}_\mathrm{I},
\end{equation}
which can be unitarily rotated back to $\ket{\psi}_\mathrm{I}$.
On the other hand, when the trajectory state is found to be $\ket{+}_\mathrm{T}$ (which happens with probability 3/4), the output state is partially mixed:
\begin{equation}\label{eq:plusPS}
\begin{pmatrix}
|\beta|^2 & -i \alpha^{*} \beta/3\\
i\alpha \beta^{*}/3 & |\alpha|^2
\end{pmatrix}.
\end{equation}
This output state has a reduced purity, but it still maintains some coherence.
Although this is not necessarily the optimal measurement strategy or the best noise model to showcase this scheme,
it illustrates that a quantum-controlled superposition of noisy channels allows some coherence to reach the receiver.
Hence, the sender and the receiver can communicate some quantum information.
We will quantify the amount of quantum information precisely in Section \ref{sec:experiment}.
Communication advantages in this case have been attributed to the ability to quantum coherently control which channel to use \cite{abbott2018}.

This type of architecture is relatively easy to imagine deploying in practice.
Most modern quantum communication takes place via optical fibers. As is often the case, these fibers can be noisy, resulting in a reduced ability to transmit information. 
Since a photon can easily be sent through a superposition of two (or more) fibers, the use of such parallel architectures could already improve security in existing communication networks.

\subsection{Channels in Series with Quantum-Controlled Operations}
\label{subsec:seriesQCO}

A different way to significantly reduce the noise produced by some channels is to let them be traversed by two trajectories in a superposition, and by allowing different operations in each branch of the superposition. In this case, we will place our two channels in series, resulting in the architecture presented in Fig.~\ref{cartoon}\textbf{b)}.
In each branch, the channels 1 and 2 are placed in the same order, and different unitary operations may be inserted. (Such unitary operations are labeled as $U_1$, $U_2$ and $U_3$ in Fig.~\ref{cartoon}\textbf{b)}. In principle, however, more operations could be inserted along the trajectories).
This scheme was originally presented in Ref.~\cite{guerin2019}, where it was referred to as a `superposition of direct pure processes'.

Let us now consider the action of the superposition of trajectories in series with the noise model of Eq.~\eqref{eq:XYsimp}, setting, following the notation of Fig.~1\textbf{b)}, $U_1=Y$, $U_2=\mathcal{I}$, $U_3=\mathcal{I}$ ($\mathcal{I}$ being the identity operator).
We will again consider the initial state of the system to be $\ket{\psi}_\mathrm{I}\ket{+}_\mathrm{T}$.
This time, we will imagine performing a measurement in the computational basis on the qubit stored in the internal DOF.
As we show in Section \ref{subsec:calc}, finding the internal qubit in $\ket{0}_\mathrm{I}$ projects the trajectory state into $\ket{\psi}_\mathrm{T}$, while finding it in $\ket{1}_\mathrm{I}$ projects the trajectory state into $X\ket{\psi}_\mathrm{T}$.
Hence, this superposition of trajectories perfectly filters out the noise arising from the noisy channels. (Notice that the ability to completely restore an arbitrary initial state of the information qubit implies that, were the information carrier initially entangled with an additional qubit, due to linearity this entangled state would be completely restored in turn.)

It is easy to imagine the implementation of this scheme in a real-world scenario.
The two paths (\textit{e.g.}, optical fibers) are simply sent through a few noisy transmission channels in series. (For the scheme to work, the action of each noisy channel must be correlated along the different paths.)
Since the two paths are physically distinct, the different unitary operations can easily be applied in each branch of the superposition independently. Such operations can be performed with
simple linear optical elements, 
or even directly using calibrated optical fibers, which always implement some unitary polarization rotation.
In Ref.~\cite{guerin2019} it was also pointed out that, by superposing more than two trajectories, one can perfectly compensate for any arbitrary noise.

\subsection{Quantum-Control of Channel Order}
\label{subsec:QCCO}

The original source of inspiration for this architecture is the quantum switch \cite{Chiribella2013}, a higher-order operation which takes quantum gates and applies them in a quantum superposition of  alternative orders.
Within quantum-interferometry, a quantum-optical switch exploiting superposition of trajectories in flat space-time has been proposed \cite{PhysRevLett.113.250402, PhysRevA.89.030303, private_communication}, and experimentally demonstrated \cite{procopio2015,rubino2017,rubino2017experimental,goswami2018,goswami2018communicating,wei2019experimental,guo2020experimental,Taddei2020}.
For two quantum operations, this is a quantum process in which a particle is placed in a superposition of two paths, each of which is routed through the two quantum operations in alternative orders (see Fig.~\ref{cartoon}\textbf{c)}).
This scheme features all the necessary requirements for an advantage in quantum information processing over standard channels \cite{guo2020experimental,goswami2018communicating}, and it can be provably characterized
as a causally-indefinite process \cite{rubino2017,rubino2017experimental,goswami2018, Ara_jo_2015, Oreshkov_2016,Branciard2016,oreshkov2019}. 

Applying the quantum switch to two copies of the channel in Eq.~\eqref{eq:XYsimp}, one finds that the output state is \cite{chiribella2018indefinite}
\begin{equation}\label{eq:outputSWITCH}
\frac{1}{2}\ket{\psi}_\mathrm{I}\bra{\psi}_\mathrm{I}\otimes\ket{+}_\mathrm{T}\bra{+}_\mathrm{T} + \frac{1}{2} Z\ket{\psi}_\mathrm{I}\bra{\psi}_\mathrm{I}Z\otimes \ket{-}_\mathrm{T}\bra{-}_\mathrm{T}.
\end{equation}
Analogously to the previous two examples, we will now measure the trajectory in the $\{\ket{+}_\mathrm{T},\ket{-}_\mathrm{T}\}$ basis.
If the outcome is $\ket{+}_\mathrm{T}$, the state has been transmitted perfectly, whereas if one finds $\ket{-}_\mathrm{T}$, a simple phase correction is required to exactly restore the initial state.

The resources required to implement the quantum-optical switch in the laboratory are relatively minimal, it simply requires linear optical elements to route the photon through the two noisy channels in a superposition of their orders.
However, in order to be effective, this layout requires the action of the two noisy channels on the photon to be suitably correlated both in space and time (as in Fig.~3 of Ref.~\cite{kristjansson2020single}). Instead, in standard communication networks \cite{PhysRevA.77.052325}, the noisy regions are usually localized in space and fixed in time. Any such network would thus require the photon to travel back toward the sender to enter the second channel, and this scheme requires  this return trip to occur without traversing any further noisy region (which could happen if the two channels introduce noise in the direction from the sender to the receiver, but not vice versa).

\subsection{Comparison}

\emph{Comparison of different schemes in this work ---}
In all of the three schemes above, a particle is placed in a quantum superposition of two trajectories which are then routed through  various devices and noisy communication channels.
All three methods result in a coupling of the internal state to the state of the trajectory, and the two trajectories must maintain coherence in order to show a communication advantage.
(Practically, this is required since the trajectory is measured in a superposition basis.)

This coupling to the trajectory DOF is a necessary requirement to achieve any advantages. 
In the parallel and indefinite order layouts the channels themselves or the routing through their different orders give rise to the coupling, whereas in the series scheme this coupling is created by the quantum-controlled operations. In light of this, it has been proposed that the quantum-controlled operations used in the superpositions of channels in series (Fig.~\ref{cartoon}\textbf{b)}) should be considered as additional resources (referred to as `encoding', `decoding' and `repeaters' in Refs.~\cite{kristjansson2019,chiribella2019quantum}), as they can couple the internal DOF to the trajectory independently of the choice of noisy channels. However,
these operations do not require any additional experimental resources beyond the transmission lines themselves (for example, polarization rotations can arise from the mere twisting of optical fibers, and are effectively unavoidable), which are the same experimental resources used for the other two schemes.

It was also noted that the number of noisy channels traversed by the particle in each branch of the superposition differs between the three schemes~\cite{abbott2018,kristjansson2019,Grinbaum2020,kristjansson2020single}: the quantum-control of parallel channels contains only one channel in each interferometer arm, whereas the other two schemes contain two channels per arm. 
%
When the information carrier crosses several noisy channels in sequence, the overall noise is always equal to (in the case, \textit{e.g.}, of two EB channels) or greater than (\textit{e.g.}, in the case of two depolarising channels of the form $\rho' = p \rho + (1-p) \frac{\mathcal{I}}{2}$) that introduced by one channel.
However, although the quantum-control of channel order needs at least two channels to create the required coupling between the trajectory and the internal DOF, it is still able to overcome the (potentially additional) noise caused by the multiple noisy channels.
\emph{Comparison to previous work ---} The origin of the communication enhancement in the three schemes studied here has been a subject of recent debate in the literature~\cite{abbott2018,kristjansson2019,chiribella2019quantum,guerin2019,Grinbaum2020}. This debate revolved around the understanding of the role of causal indefiniteness in the task of noise reduction. In fact, after it was discovered that such an enhancement could be achieved by placing the channels in an indefinite causal order, it was later found that other configurations, which did not have an indefinite causal order, could achieve the same or even a better enhancement. This called into question whether indefinite causality is necessary to achieve such effects.

In this work, we compare experimentally and numerically all the proposed setups leading to noise cancellation in quantum communication, and we quantify the achieved advantages over quantum communication schemes with classical trajectories.
This represents the first experimental comparison among these different schemes, which provides an answer to the debate on experimental grounds by presenting an experimentally-relevant analysis of the resources in play.
In particular, we compare the schemes with respect to their  experimental requirements within an interferometric paradigm, rather than studying them as higher-order operations from a strictly theoretical viewpoint~\cite{kristjansson2019,chiribella2019quantum}.
To this end, we focus on the following four points: \textit{i.} we illustrate that all three schemes use the same resource when considering experimental quantum interferometry, \textit{ii.} we show that this resource is the coupling of the degree of freedom carrying the information to the trajectory degree of freedom, \textit{iii.} we experimentally prove that, for the set of tested noisy channels, the superposition of channels in series with quantum-controlled operations features the highest performance, and \textit{iv.} we numerically show that, in the vast majority of cases, this holds for generic randomly-generated channels.

We will now proceed to quantify the amount of quantum information which can be transmitted using these various schemes individually for different noise models.
Overall, we find that, for all the types of noise considered,  the use of channels in series with quantum-controlled operations exceeds or equals the performance of the quantum-control of parallel channels and quantum-control of channel order.

\section{Quantifying Channel Performance}
\label{sec:Quantif_Channel_Performance}

In order to rigorously compare the ability of  the different schemes to transmit quantum information, an experimentally accessible figure of merit is necessary.
The quantum capacity $\mathcal{Q}(\cal C)$ of a channel $\cal C$ is the number of qubits that are transmitted for each use of that channel \cite{Lloyd1997, Devetak2005}. 
In general, this is a rather complex function that can be difficult to even theoretically assess, making its use as a quantifier somewhat limited in practice. However, it is lower bounded by \cite{Lloyd1997}:
\begin{equation}
\mathcal{Q}(\mathcal{C})  \ge \max\limits_{\rho_{AB}} \, \mathcal{I}_{\mathrm{c}}({\mathcal{C}},\rho_{AB}),
\end{equation}
where  $\mathcal{I}_{\mathrm{c}}$ is the coherent information  \cite{Schumacher1996} of the channel with respect to $\rho_{AB}$, which is defined as
\begin{equation}\label{eq:Ic}
\mathcal{I}_{\mathrm{c}}({\mathcal{C}},\rho_{AB}) : = S   ( \rho_B' )-  S(\rho_{AB}') \, ,
\end{equation}
where $\rho_{AB}$ is a bipartite state, $\rho_{AB}'  : =  (\mathcal{I}_A \otimes \mathcal{C})  (\rho_{AB})$ is the output state obtained by applying channel $\mathcal{C}$ on system $B$,  $\rho_{B}'  := {\mathrm Tr}_A  [\rho_{AB}']$ is its marginal state, and $S(\rho)  := - {\rm Tr} [\rho \log \rho]$ is the von Neumann entropy.
Although a comparison of the coherent information of two channels does not necessarily translate into a comparison of the quantum capacity of the channels (except, of course, when the lower bound is maximal, as in this case it coincides with the quantum capacity), we will employ it as our quantifier for channel performance here. In addition to the practical motivation of being a more readily computable quantifier, this choice is further motivated by the fact that the coherent information has an operational meaning beyond that of the quantum capacity. Namely, it provides the one-way distillable entanglement when maximised over local operations performed by the sender, and, if maximised over LOCC operations, it gives the two-way distillable entanglement \cite{Devetak2005_Distillation}, which is typically considered to be the amount of `useful entanglement' which can be transmitted using the channel.

Let us briefly consider a few simple examples, assuming a 2-qubit system, with $\rho_{AB}$ set to a maximally-entangled pure state. 
If the channel is unitary (\textit{i.e.}, noiseless), then the initially pure state remains pure after the application of the channel, thus $S\bigl({\cal C}(\rho_{AB}')\bigr)=0$.
However, since $\rho_{AB}$ is maximally entangled, tracing out the subsystem $A$ will leave the subsytem $B$ in a maximally mixed one-qubit state with entropy $S\bigl({\cal C}(\rho_{B}')\bigr)=1$.
Therefore the coherent information of a unitary channel, with respect to a maximally entangled probe state is $1$.
If, on the other hand, the channel induces decoherence, the entangled probe state will become mixed, and the second term will increase: $S\bigl({\cal C}(\rho_{AB}')\bigr)>0$. 
Because the first term cannot be larger than $1$, as decoherence is induced the coherent information decreases.
$\mathcal{I}_{\rm c}({\cal C},\rho_{AB})$ is often maximized when $\rho_{AB}$ is a maximally-entangled state.
This was proven to be the case for the quantum switch and a specific noisy model in \cite{salek2018quantum}.
Furthermore, numerical optimisations suggest that a maximally-entangled Bell state maximises the coherent information for the cases we study here.
In any case, the evaluation of $\mathcal{I}_c$ for any arbitrary state sets a lower bound for the quantum capacity of the channel.
Throughout the rest of this manuscript, when we refer to the coherent information, we do so with reference to a maximally-entangled Bell state. 

One could consider estimating $\mathcal{I}_c({\cal C},\rho_{AB})$ directly by probing the channel with an entangled state \cite{cuevas2017}. In this case, however, the trade-off is that this state will be more prone to errors in the preparation phase (and such errors are to be considered in addition to all the others already mentioned).
Thus, our experimental approach will be to first perform quantum process tomography on the superposition of communication channels.
With the resulting estimate of the experimental channels, we will then be able to compute $\mathcal{I}_c$ with ideal maximally-entangled states, and will use this metric to quantify the performance of the various schemes.

In our experiment, we study single-qubit channels acting on the polarization DOF, and equal superpositions of trajectories.
Since, as we have seen previously, the coherence between the two trajectories is crucial, both the internal DOF and the trajectory must be fully characterized. 
In general, this requires two-qubit process tomography on the path (trajectory) qubit and the polarization qubit.
To perform this characterization, we use heralded single photons in order to maintain the connection to the interpretation of the quantum capacity as the information transmitted per information-carrying system.

Notice that the sender only ever encodes information in the polarization DOF, whereas the receiver must measure both the trajectory and the polarization DOFs. Hence, this is effectively a 1-to-2 qubit channel. Because of this, performing full two-qubit process tomography provides more information than is strictly required.

\section {Experiment}
\label{sec:experiment}

\begin{figure*}
  \centering%
      \includegraphics[width=\textwidth]{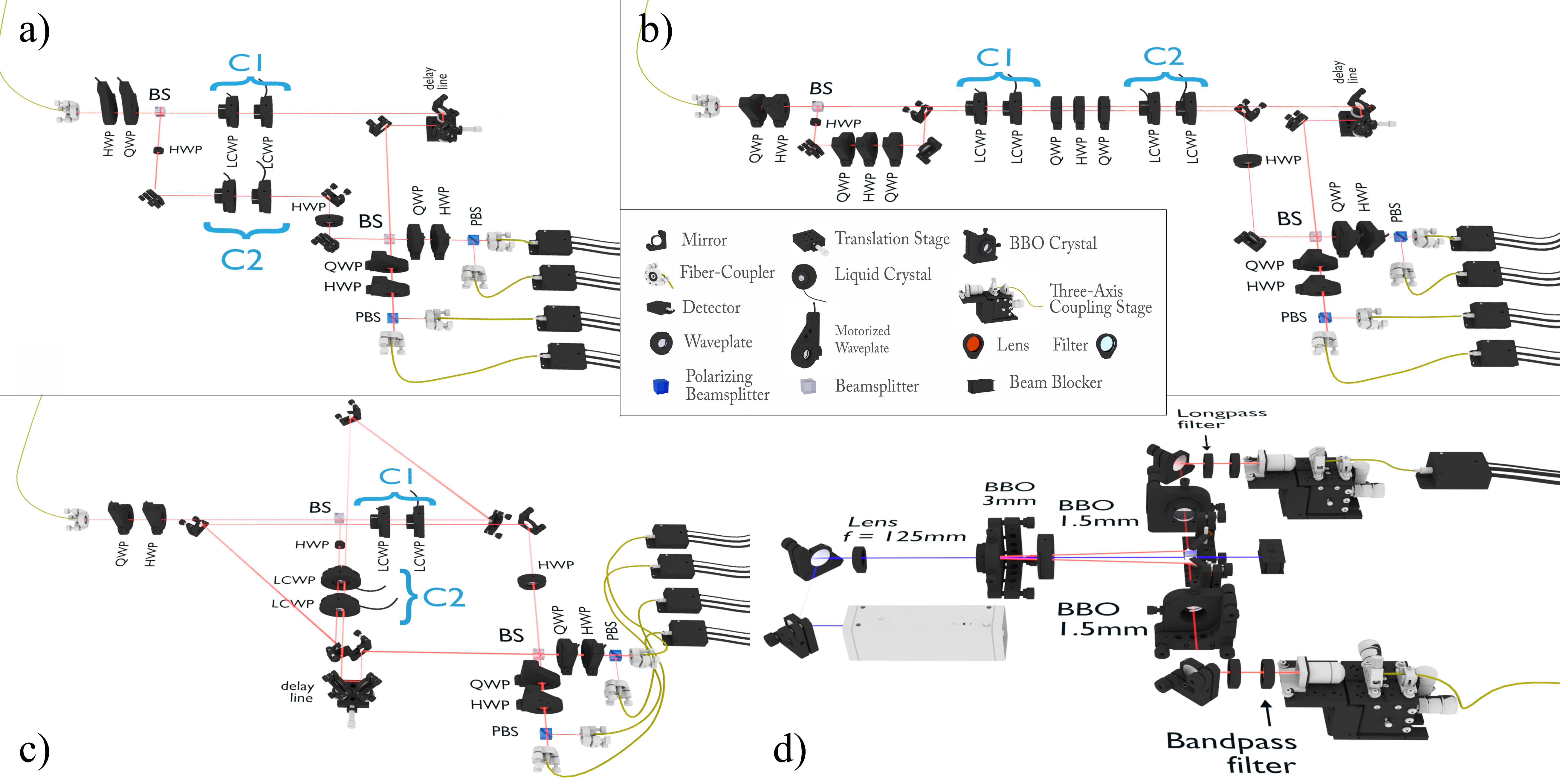}
  \caption{\textbf{Experimental Setup.} \textbf{a)} \textbf{Quantum-Control of Parallel Channels.} After their polarization is set via a half waveplate (HWP) and a quarter waveplate (QWP), single photons are injected into a Mach-Zehnder interferometer. One noisy channel is placed into each arm of the interferometer, and each channel is realized through two liquid crystal waveplates (LCWP), the first positioned at 0$^\circ$ (to implement $\mathcal{I}$ or $Z$ by changing the retardance), the second at 22.5$^\circ$  ($\mathcal{I}$ or $X$). By means of a piezo-electric trombone delay line, the photon interfering on the second beamsplitter of the interferometer can be projected onto the bases $\{\vert+\rangle_\mathrm{T},\vert-\rangle_\mathrm{T}\}$ or $\{\vert R\rangle_\mathrm{T},\vert L\rangle_\mathrm{T}\}$ of the trajectory. Finally, the photons' polarization is measured through a sequence of QWP, HWP and a polarizing beamsplitter. \textbf{b)}  \textbf{Channels in Series with Quantum-Controlled Operations.} As in the previous scheme, the photons are prepared in polarization via QWP and HWP and injected into a Mach-Zehnder interferometer. In this case, the two noisy channels are placed in the two superposed branches in series with the same order. Also in this case, the channels are realized through LCWPs. Furthermore, before each noisy channel, additional unitary operations are realized through sequences of QWP, HWP and QWP (before the first channel, the QWP, HWP and QWP are placed in one branch of the trajectory only, whereas between the two channels the waveplates are in both branches, since we only implement cases where $U_2 = U_3$). The rest of the setup is the same as in the previous case. \textbf{c)} \textbf{Quantum-Control of Channel Order.}  The preparation and measurement of the photons in polarization happens as in the previous schemes, as well as the realization of the noisy channels, and the projection of the trajectory DOF. In this case, however, the Mach-Zehnder interferometer is folded into two loops so that the photon can travel through the two channels in the two alternative orders in each arm of the interferometer. \textbf{d)} \textbf{Heralded single-photon source.} We generate photon pairs using a type-II spontaneous-parametric-down-conversion source. One photon is directly detected with an avalanche photodiode (upper arm), whereas the other is coupled into an optical fiber and sent to one of the setups \textbf{a)}, \textbf{b)} or \textbf{c)}. The interferometers in setups \textbf{a)}, \textbf{b)}, and \textbf{c)} all contain two compensation HWP at the beginning and at the end of the reflected arm, so as to compensate for the phase shifts due to the reflection from the beamsplitter.}
\label{setup}
\end{figure*}

\emph{Implementing Noisy Channels---} In our experiment, we encode and transmit information in the internal polarization DOF.
We induce noise on this DOF using liquid crystal waveplates (LCWP), which can rapidly implement different polarization rotations to effectively decohere the polarization state in a precise and controllable manner \cite{rozema2014optimizing}.
The LCWP retardance can be changed between $0$ rad and $2\pi$ rad in approximately 100 ms by varying the applied voltage (see Section \ref{subsec:LiquidCrystals} for more details). Using these fast LCWPs we can change the operations on-the-fly to actively decohere the photon's polarization, in contrast to previous experiments wherein decoherence was achieved by averaging the results during the data analysis \cite{goswami2018communicating, guo2020experimental}.  Nevertheless, the two methods yield the same results, so we will make use of both techniques interchangeably.
Physically, the noise models we study can be understood as randomly applying one of four operations ($\mathcal{I}$, $X$, $Y$, or $Z$). The probability of each operation to occur defines the noisy channel.  

Specifically, we implement four different noisy channels. The first is a generalization of the entanglement-breaking channel ${\cal C}_{\text{EB}}(\rho)$ discussed above. However, in this general case, the $X$ and $Y$ operations are applied with probability $1-p$ and $p$, respectively (one recovers the ${\cal C}_{\text{EB}}(\rho)$ for $p=1/2$):
\begin{equation}\label{XY}
{\cal C}_{XY}^p(\rho)= (1-p) X\rho X + p Y\rho Y.
\end{equation}
We also study the well-known bit-flip (BF) ${\cal C}_{\text{BF}}^p(\rho)$ and phase-flip (PF) ${\cal C}_{\text{PF}}^p(\rho)$ (or dephasing) channels:
\begin{subequations}
\begin{align}\label{bitFlip}
{\cal C}_{\text{BF}}^p(\rho)&= (1-p) \, \rho  + p X\rho X,\\
\label{dephase}
{\cal C}_{\text{PF}}^p(\rho)&= (1-p) \, \rho  + p Z\rho Z,
\end{align}
\end{subequations}
respectively.
Finally we study a depolarizing channel ${\cal C}_{\text{BB84}}^p(\rho)$, known as the BB84-channel \cite{BB84channel}:
\begin{align}\label{BB84}
{\cal C}_{\text{BB84}}^p(\rho)=& \, (1-p)^2 \rho  + (1-p) \, p \, X\rho X \notag\\
&+ (1-p) \, p \, Z\rho Z + p^2 \, Y\rho Y.
\end{align}
For the BB84-channel, when the noise probability is $p=0.5$ the channel is completely depolarizing, mapping any input to the maximally-mixed state. In Section~\ref{subsec:Random_Channels}, we also report a numerical estimation of the performance of the three layouts in the generic case of randomly-generated channels.

To realize a single channel we use two LCWPs.
The first LCWP's optic axis is set to $0^\circ$, and can thus implement either $Z$ or the identity operation by setting the retardance to $\pi$ rad or $0$ rad, respectively.
The second LCWP's optic axis is set to 22.5$^\circ$ to execute $X$ or the identity operation, again by setting the retardance to $\pi$ rad or $0$ rad, respectively.
When the first LCWP performs $Z$ and the second $X$, the net result is $Y$ (up to a phase).
Hence, with these two LCWPs we are able to carry out all four required unitary operations, and switch between them in about $100$ ms.

In light of this, a straightforward implementation would be to generate a random number from some defined distribution before a photon enters the channel, and then to set the operations accordingly.
However, the net result is the same if we allow several photons to pass through the channel for each coin flip, provided that we average over a sufficiently large number of coin flips.
This is advantageous, as it allows us to increase the single-photon count rate well above the switching speed of the LCWPs.
In our experiment, we employ two different methods for the data acquisition.
In the first, we change the applied operation every second. Since our photon rate (detected at the output, after the experiment) is of the order of $3000$ Hz, this means that approximately $3000$ subsequent heralded photons experience the same unitary operation (see Section \ref{subsec:Data_Acquisition} for more details). 
Our Monte Carlo simulations show that, with these numbers, 100 seconds (and 100 different operations) per measurement setting are sufficient to achieve a process fidelity (\textit{i.e.}, the fidelity to achieve the desired noisy channel) above 99\% (for details, see Section \ref{subsec:Process_Fidelity}).
In order to ensure an optimal implementation while maintaining a reasonable duration of the data-taking procedure, we used 1000 different internal configurations for our experiment, resulting in a fidelity of 99.98\% per channel.
In the second technique, we simply take data for each input state and each measurement setting with the LCWPs set to implement a fixed unitary operation.
We then weight the data from these different configurations according to the probability distribution of the desired noise model. (This method was also demonstrated in \cite{goswami2018communicating, guo2020experimental}.) 

\vspace{3mm}

\emph{Creating Superpositions of Trajectories---} As shown in Fig.~\ref{setup}, we experimentally create different superpositions of trajectories by placing single photons in an equal quantum superposition of paths using a 50/50 beamsplitter.
The single photons are generated with a standard type-II down-conversion source described in Fig.~\ref{setup}\textbf{d)} and in Section \ref{subsec:Source}.
These two paths (trajectories) are then routed through a series of LCWPs, which implement different noisy channels, in a parallel configuration (Panel \textbf{a)}), in series (Panel \textbf{b)}), or in a quantum superposition of the two alternative orders (Panel \textbf{c)}).

All three set-ups are realised through Mach-Zehnder interferometers. In the first case (Fig.~\ref{setup}\textbf{a)}), one channel is placed in each interferometer arm. In the second case (Fig.~\ref{setup}\textbf{b)}), the channels are arranged in series in both arms of the interferometer, and additional operations are performed before each channel through waveplates. Finally, the third scheme (Fig.~\ref{setup}\textbf{c)}) is accomplished using a folded Mach-Zehnder interferometer in which the two channels appear in alternating order in each of the interferometer's arms.
The setup presented in Fig.~\ref{setup}\textbf{c)} represents a possible realization of a quantum-optical switch wherein the system qubit is encoded in the polarization DOF, and the control qubit in the path DOF.
Other encodings for this type of process have been proposed \cite{rambo2016functional, PhysRevA.89.030303} and experimentally demonstrated \cite{goswami2018, wei2019experimental}. 
Regardless of the detailed implementation, all proposals to implement a quantum-optical switch use one DOF to route a photon through channels in different orders, while the channels act on some other DOF.

In order to perform quantum process tomography to extract the coherent information, we must prepare a tomographically-complete set of input states, and measure in a tomographically-complete number of different bases.
In brief, we use waveplates before the first beamsplitter to prepare the state of the polarization qubit in either $\ket{0}_\mathrm{I}$, $\ket{+}_\mathrm{I}$, $\ket{R}_\mathrm{I}$, or $\ket{L}_\mathrm{I}$ (where $\ket{R}_\mathrm{I}=(\ket{0}_\mathrm{I}-i\ket{1}_\mathrm{I})/\sqrt{2}$, $\ket{L}_\mathrm{I}=(\ket{0}_\mathrm{I}+i\ket{1}_\mathrm{I})/\sqrt{2}$), and waveplates and polarizing beamsplitters after the second beamsplitter to measure in all bases (\textit{i.e.}, $\{ \ket{0}_\mathrm{I},\ket{1}_\mathrm{I}\}$, $\{ \ket{+}_\mathrm{I},\ket{-}_\mathrm{I}\}$, and $\{ \ket{R}_\mathrm{I},\ket{L}_\mathrm{I}\}$).
In our experiment, we set the state of the path DOF to $\ket{+}_\mathrm{T}$, $\ket{-}_\mathrm{T}$, $\ket{R}_\mathrm{T}$, or $\ket{L}_\mathrm{T}$ by varying the relative phase of the paths after the first beamsplitter using a pair of mirrors placed on a trombone-delay stage controlled by a piezo-electric actuator.
To prepare the state to $\ket{0}_\mathrm{T}$ or $\ket{1}_\mathrm{T}$, we simply block one or the other path.
We measure the path DOF analogously, by setting the different phases, or blocking one of the two paths.
The full details of our process tomography protocol are presented in Section \ref{subsec:process_tomography}.

\section{Results}
\label{sec:Results}

\begin{figure}
  \centering%
      \includegraphics[width=\columnwidth]{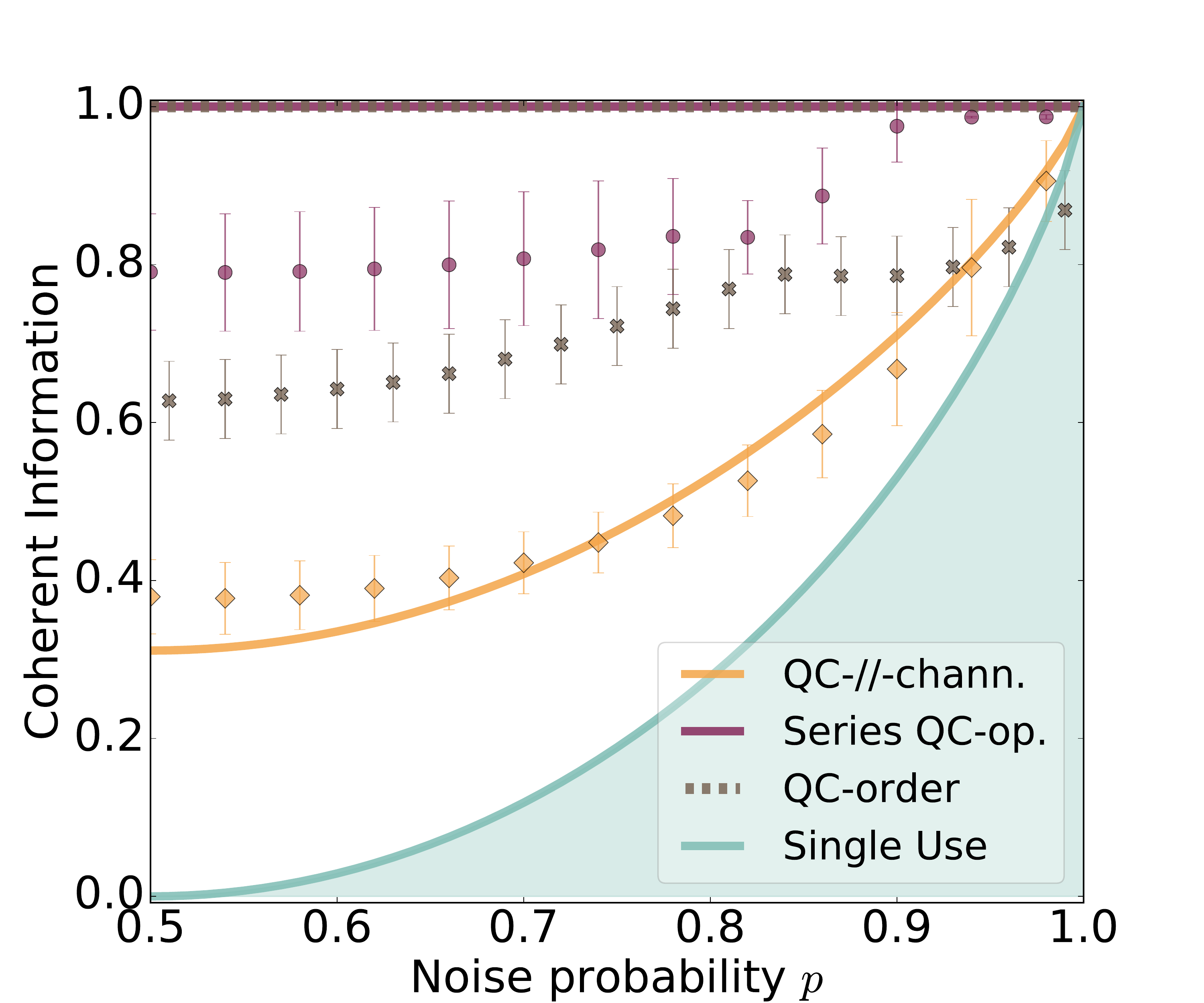}
  \caption{\textbf{Experimental \textit{XY}-Channel noise data.} The theoretical trends associated with the channels in series with quantum-controlled operations and the quantum-control of channel order show full activation. The experimental data do not perfectly match the theoretical trends because, for $p=0.5$, the channel produces an equal mixture of $X$- and $Y$-operations, and such case can be experimentally realised with a lower fidelity than the one in which only one of the two operations is performed (\textit{i.e.}, when $p=0$ or 1). It follows that, in the central region, the experimental data are further apart from the theoretical trend than they are on the upper end. The quantum-control of parallel channels does not allow full activation, and thus it is positioned below the previous two trends. In this case, the experimental data are closer to the theoretical expectation. The reason of the higher agreement is that, in the case of the disposition of noisy channels in parallel, only one channel is present in each branch of the interferometer. As a consequence, the experimental imperfections affecting each branch are smaller than in the dispositions of channels in series and in indefinite order. Finally, the coherent information associated to only one $XY$-channel is theoretically lower than all the other layouts. A detailed analysis of the error estimation and the systematic error is provided in Section \ref{subsec:Data_Acquisition}. The labels `QC-//-chann.', `Series QC-op.' and `QC-order' stand for `quantum-control of parallel channels', `channels in series with quantum-controlled operations' and `quantum-control of channel order', respectively. The same labels will be used in all plots.}
    \label{fig:XYchan}
\end{figure}

\begin{figure}
  \centering%
      \includegraphics[width=\columnwidth]{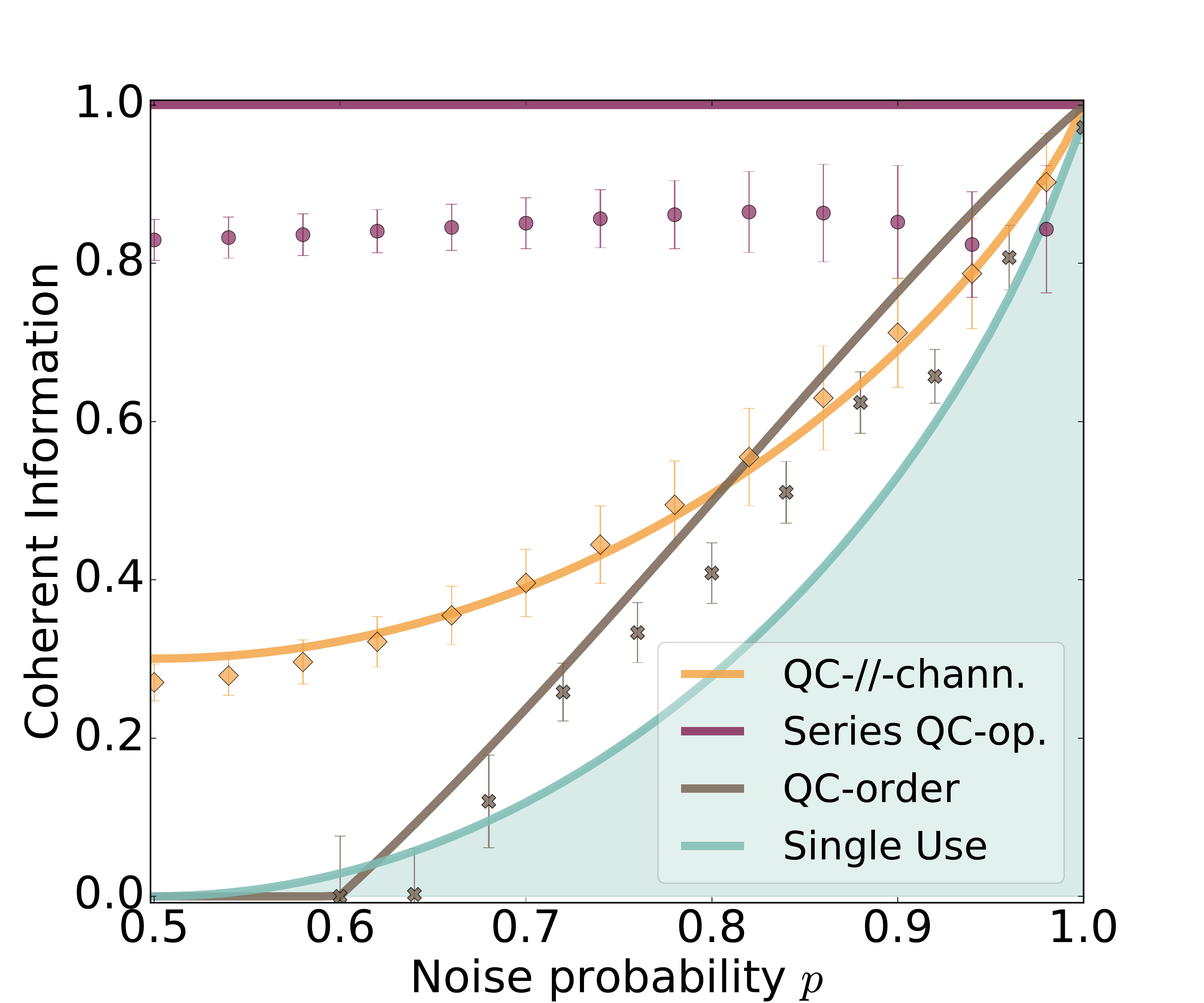}
  \caption{\textbf{Experimental BF- and PF-noise data.} The experimental data of quantum-control of parallel channels and the quantum-control of channel order are in good agreement with the theoretical trends. Conversely, the configuration of the channels in series with quantum-controlled operations shows a constant offset between the experimental data and the expected theoretical trend. This discrepancy is due to the fact that, in this case, all the liquid crystals are arranged in series, with the additional presence of waveplates realizing a Hadamard gate, and hence this configuration is the one that exhibits the greatest amount of experimental imperfections along each path.  In spite of this, for most values of $p$ the coherent information that can be achieved with the series configuration is still above all others by several standard deviations.}
    \label{fig:BitPhasechan}
\end{figure}

\begin{figure}
  \centering%
      \includegraphics[width=\columnwidth]{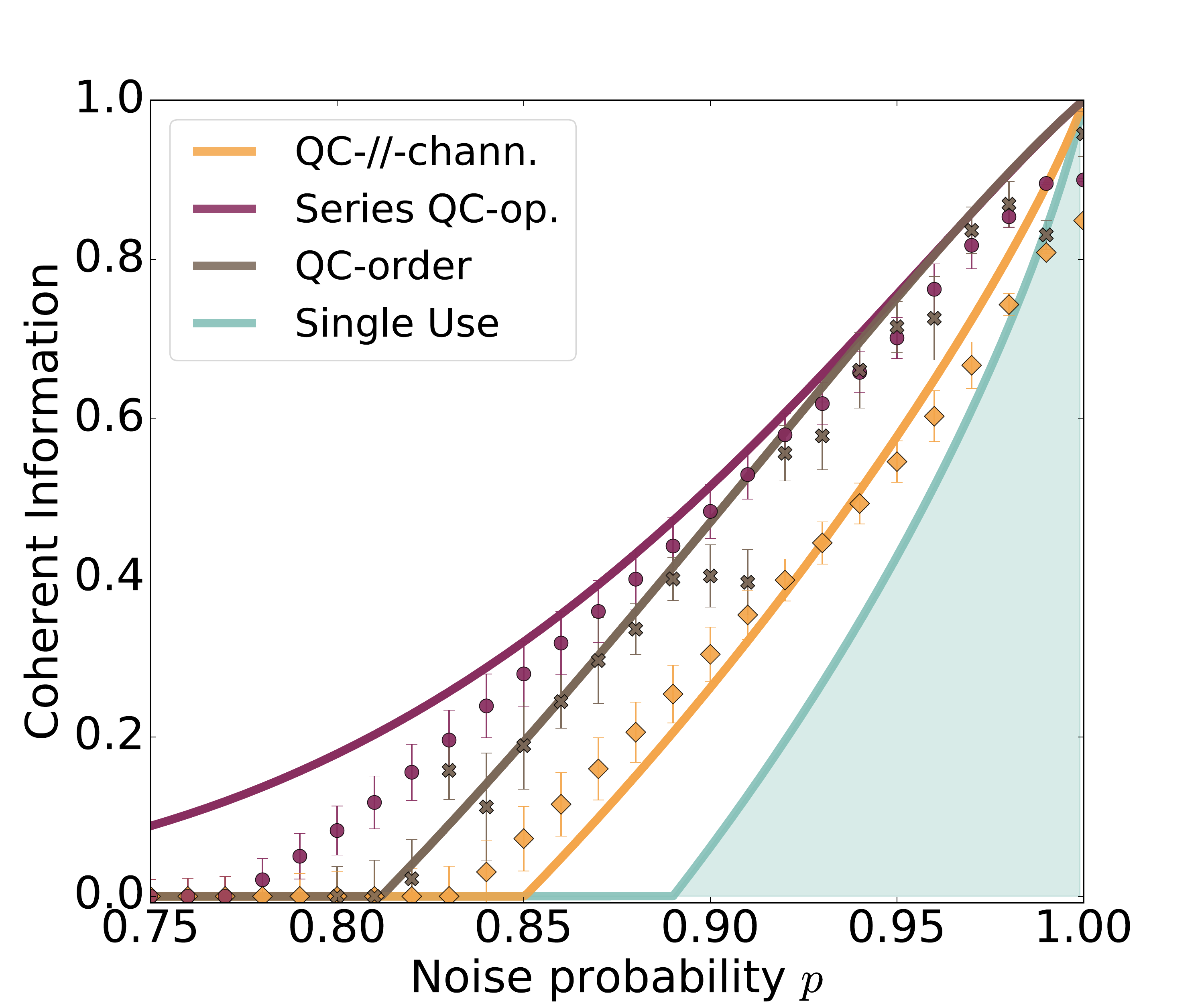}
  \caption{\textbf{Experimental BB84-channel noise data.} As in the previous plots, the continuous lines show the expected theoretical trends, while the squares, circles and crosses represent the experimental data corresponding to the quantum-control of parallel channels, the channels in series with quantum-controlled operations, and the quantum-control of channel order, respectively. All the experimental data are in high agreement with the expected theoretical trends.}
    \label{fig:BB84}
\end{figure}

\emph{XY-Channel---} Below, we present our results for the three combinations of the noisy channels described in equations \eqref{XY}--\eqref{BB84}.
We will first consider two copies of the \textit{XY}-channel (Eq.~\eqref{XY}). 
In Section \ref{sec:QSuperp_Trajectories}, we observed that when $p = 0.5$ both the channels in series with quantum-controlled operations and the quantum-control of channel order (with $U_1=Y$, $U_2=U_3=\mathcal{I}$) are able to transmit quantum information perfectly. In Fig.~\ref{fig:XYchan}, we observe that such a perfect `activation' (in our case, the term refers to a combination of two noisy channels which enables one to communicate through such a combination with less noise than either individually) is theoretically possible for all values of $p$.
In fact, the purple and brown lines show the coherent information for two \textit{XY}-channels combined in series and in indefinite order, respectively.
For both of these situations, the theoretical coherent information is equal to $1$ for all $p$, meaning that one qubit per use can be transferred.
In the same plot, our experimental data are presented as squares (for the quantum-control of parallel channels), circles (for the channels in series with quantum-controlled operations) and crosses (for the quantum-control of channel order) with matching colors.
The dominant source of the statistical errors is the uncertainty in determining the initial states for the process tomography. (In fact, the input states were prepared and characterized at the output of the source, but they were then sent to each experiment via 3m long optical fibers, which introduced additional noise.) This uncertainty leads to the error bars on all the data sets presented in Figs.~\ref{fig:XYchan}--\ref{fig:BB84} (see Section \ref{subsec:Data_Acquisition} for more details on error estimation).
Nevertheless, all plots display a good agreement between experiment and theory. 
As expected, the experimentally measured coherent information is slightly lower than that predicted theoretically.
This offset is mainly due to the following systematic errors: \textit{i.}\ the imperfect visibility when the two trajectories are recombined on the second beamsplitter, \textit{ii.}\ phase drifts which can occur during the experimental runs, and \textit{iii.}\ slight calibration errors in the LCWPs implementing the channels and the waveplates used for state preparation and measurement.
These systematic effects are not included in the calculation of our experimental errors.
Full details of the measurement procedure, including photon count rates and measurement times, as well as the statistical and systematic errors affecting the data are presented in Section \ref{subsec:Data_Acquisition}.

The orange data set reported in Fig.~\ref{fig:XYchan} corresponds to the coherent information when the two \textit{XY}-channels are used in a quantum-controlled superposition.
In Section \ref{sec:QSuperp_Trajectories}, we illustrated that, when $p=0.5$, the output still displays a partial dependence on the input state.  
However, calculating the coherent information reveals that this is not sufficient to transmit a single qubit per use (\textit{i.e.}, the coherent information is less than $1$).
Nevertheless, the orange curve indicates that quantum information can still be transmitted, although not at the maximum rate.

The turquoise curve in Fig.~\ref{fig:XYchan} represents the coherent information of a single trajectory traversing a single copy of the channel, which is $1-H(p)$, where $H(p) = -p \, \mathrm{log}(p)-(1-p)\,\mathrm{log}(1-p)$ is the Shannon entropy. (The shaded area underneath represents the region within which any activation by either channel layout is less effective than directly using one of the noisy channels.) Because in our experiment we assume that the noise strengths $p$ of the two channels are always identical, using the channels in a classical mixture, as depicted in Fig \ref{cartoon}\textbf{d)}, will also result in the capacity of a single use of the channel.
If a single trajectory was sent through two copies of the channel in a row, the coherent information would be even lower, since the second channel would further decohere the polarization state.
We see in this first case that for all values of the noise parameter $p$, all three superposition methods transmit more quantum information than only using a single-trajectory.

\vspace{3mm}

\emph{Bit-Flip and Phase-Flip Channels---} Ref.~\cite{salek2018quantum} showed that a quantum superposition of the causal order of a bit-flip and a phase-flip channel can transmit more quantum information than the amount which can travel through each channel individually.
(Referring to Fig.~\ref{cartoon}, this corresponds to replacing channel 1 with the bit-flip channel (Eq.~\eqref{bitFlip}), and channel 2 with the phase-flip channel (Eq.~\eqref{dephase}). Note that, contrary to the other cases, here we consider two different types of noisy channels C1, C2, rather than two copies of the same channel).
In light of this, Ref.~\cite{guerin2019} pointed out that this idea can also be applied when the noisy channels are placed in series, provided that one allows quantum-controlled operations before and between them, and that this trick allows one to transmit quantum information perfectly (when $U_1=Y$, $U_2=U_3=H$, where $H$ is the Hadamard operation).

We experimentally confirm the predictions of Refs.~\cite{salek2018quantum, guerin2019} in Fig.~\ref{fig:BitPhasechan}.
There, we see that, regardless of the noise strength, the channels in series with quantum-controlled operations can, in theory, perfectly transmit quantum information (\textit{i.e.}, the purple line is equal to $1$).
Our experimental data (purple circles) confirm this, although they do show a slight offset due to the systematic errors discussed above.
In this case, the quantum-control of channel order (brown curve for theory, and crosses for experiment) does not work as well.
Nonetheless, we do find that for a range of $p$ it outperforms the single use value $1-H(p)$.
For this choice of noisy channels, the quantum-control of parallel channels (orange curve) can transmit more information than their quantum-controlled order. 
For a large range of $p$, it is larger than the value achievable through the quantum-control of channel order, and the slight theoretical advantage of this latter over the quantum-control of parallel channels for large enough values of $p$ is not observable in our experimental data.

For a fair comparison, we mention that changing the quantum-controlled operations $U_i$ depending on the type of noise could be regarded as an additional resource. In fact, setting the optimal quantum-controlled operations requires one to characterize the noise prior to using the channels. In Section \ref{subsec:Fixing_Operations}, we compare the performance of the channels in series with quantum-controlled operations for the same unitaries that we use for the \textit{XY}- and BB84-channels (namely, for $U_1=Y$, $U_2 = U_3 = \mathcal{I}$). There, we observe that setting $U_2 = U_3 = \mathcal{I}$ results in a performance that is comparable to that of the quantum-control of channel order, and which still outperforms the single-use capacity.  In doing so, the quantum-controlled operations remain fixed in this configuration independently of the type of noise.

\vspace{3mm}

\emph{BB84-Channel---} As a final example, we consider two copies of the depolarizing BB84-channel (Eq.~\eqref{BB84}).
These results are shown in Fig.~\ref{fig:BB84}.
Also in this case, the channels in series with quantum-controlled operations (this time with $U_1=Y$, $U_2=U_3=\mathcal{I}$), shown in purple, achieves the largest enhancement.
While with only two trajectories it is not possible to perfectly transmit quantum information through these noisy channels, Ref.~\cite{guerin2019} showed that with additional trajectories any type of noise can be perfectly corrected with the quantum superposition of channels in series.
The quantum-control of channel order in this case outperforms both the single-use coherent information ($1-2H(p)$) and the coherent information of the quantum-control of parallel channels.

These three examples show that, depending on the type of noise, different superpositions of channels can lead to the ability to transmit different amounts of quantum information.
The physical origin of this ability is an effective coupling between the trajectory and the internal degree of freedom. In the present paper, this coupling is verified by the observed correlations between the states of the aforementioned two degrees of freedom. While these correlations were only sketched in the case of the EB channel in sections \ref{subsec:QCPC}-\ref{subsec:QCCO}, analogous relations hold also in the case of the other noisy channels studied in this section.
In all the cases we investigated here (wherein the schemes are used individually), even in the presence of experimental imperfections, using the channels in series with quantum-controlled operations appears to be the best candidate to evade the effects of the noise.

\section{Discussion}
\label{sec:Discussion}

In this work, we experimentally and numerically explored how the degradation of quantum information due to its propagation through noisy channels can be mitigated, and in several cases fully suppressed. This was achieved by sending quantum information carriers through a pair of noisy channels in various superpositions of trajectories. In particular, we studied three types of schemes: the quantum-control of parallel noisy channels, channels in series with quantum-controlled operations, and the quantum-control of channel order.

All of these schemes bear much in common with error filtration \cite{gisin2005}.
More recently, this has been refined in a number of theoretical works \cite{Ebler2018, salek2018quantum, chiribella2018indefinite, chiribella2019quantum, kristjansson2019, jia2019causal, procopio2020sending, caleffi2020, Sazim2020, Wilson2020, abbott2018, guerin2019, Grinbaum2020}, tied into the concept of indefinite causal orders.
While enhanced communication based on an indefinite causal order has been experimentally demonstrated \cite{goswami2018communicating, guo2020experimental}, an experimental study comparing different superpositions of trajectories in presence of various types of noise has been lacking. 
Our work is aimed at bridging this gap, by suggesting common ground based on the experimental resources that each of the analysed schemes requires. 

Our results suggest that, in most quantum-optical communication scenarios, creating a superposition of trajectories through channels in series with quantum-controlled operations should lead to the largest noise reduction.
One can easily imagine characterizing the error introduced in various communication channels, and from there setting the unitary operations accordingly. Moreover, Ref.~\cite{guerin2019} illustrated that these types of schemes can be extended to superpositions of more than two trajectories to achieve complete error cancellation for any type of noise. We have shown experimentally that with only two trajectories it is already possible to completely cancel (after accounting for experimental errors) all the noise arising from two out of the three types of noisy channels we considered. Furthermore, the quantum-controlled operations could also be introduced in the other two schemes and, potentially, they could match the performance of the layout with channels in series.

The large experimental communication enhancements presented here highlight the practical relevance of extending the quantum communication paradigm to scenarios in which not only the information carriers, but also the trajectories along which they propagate are quantum.
We expect that the relative ease of implementation of these schemes will enable them to be readily put into practice for the noise-reduction of real-world long-distance quantum communication applications.

\vspace{3mm}
\textbf{Acknowledgments:} L.\ A.\ R.\ acknowledges financial support from the Templeton World Charity Foundation (fellowship no. TWCF0194). A.\ A.\ A.\ acknowledges financial support from the Swiss National Science Foundation (SNSF) through Starting Grant DIAQ and NCCR SwissMAP. \v{C}.\ B.\ acknowledges financial support from the Austrian Science Fund (FWF) through BeyondC (F7103-N38), the project no.\ I-2906, as well as support from the European Commission via Testing  the  Large-Scale  Limit  of  Quantum  Mechanics (TEQ) (No.\ 766900) project. G.\ C.\ acknowledges financial support by the National Science Foundation of China through Grant no.\ 11675136, by Hong Kong Research Grant Council through Grants no.\ 17300918 and 17307719, and by the Croucher Foundation. H.\ K.\ acknowledges financial support from the UK Engineering and Physical Sciences Research Council (EPSRC). \v{C}.\ B., G.\ C.\ and P.\ W.\ acknowledge financial support by the John Templeton Foundation through grant 61466, The Quantum Information Structure of Spacetime (qiss.fr). P.\ W.\ acknowledges financial support from the Austrian Science Fund (FWF) through BeyondC (F7113-N38), GRIPS (P30817-N36) and NaMuG (P30067-N36), the European Commission through UNIQORN (no.\ 820474), HiPhoP (no.\ 731473) and ErBeStA (no.\ 800942), from the United States Air Force Office of Scientific Research via QAT4SECOMP (FA2386-17-1-4011), and Red Bull GmbH. P.\ W.\ and \v{C}.\ B.\ acknowledge financial support from the Foundational Questions Institute (FQXi), the Austrian Science Fund (FWF) through CoQuS (W1210-N25), and the research platform TURIS. The opinions expressed in this publication are those of the authors and do not necessarily reflect the views of the John Templeton Foundation. 
\textbf{Competing interests:} The authors declare not to have any competing interests. \textbf{Data and materials availability:} All data needed to evaluate the conclusions in the paper are present in the paper and/or the Supplementary Materials. Additional data related to this paper will be made available from the authors upon request.

\section{Appendix}
\label{sec:Methods}

\subsection{Communication Advantages when Transmitting Classical Information}
\label{subsec:classical}

The present work focuses on the transmission of quantum information through channels placed in a quantum superpositions of trajectories. This is, however, not the only possible choice: the work which initiated this research direction discussed communication advantages in transmitting classical information through a quantum-control of channel order~\cite{Ebler2018}. In the following, we briefly comment on our choice of figure of merit, and we explain how the scenarios in Fig.~\ref{cartoon} compare for transmitting classical information. 

Classical communication is determined by the amount of classical correlations which a quantum channel can maintain between its input and output. Generally, the classical capacity of a quantum channel is only zero for the class of erasure channels which replaces any input by a fixed output state. Sending quantum information, on the other hand, is a more ambitious task: here, the capability of a channel to preserve quantum correlations (\textit{i.e.}, entanglement) during the processing quantifies the transmissible quantum information. Indeed, quantum information was shown to be the most difficult to communicate~\cite{Wilde2013}, and at the same time the most valuable resource for information theoretic tasks and computation. For instance, sending quantum information can ensure secure communication~\cite{Bennett1984}, it can be used to distill secret keys for cryptography~\cite{Wilde2013}, and it is crucial for tasks like distributed quantum computation~\cite{Broadbent2009, Beals2013}. Furthermore, trivially, quantum information can also be used to communicate classical information~\cite{Hol73}. Hence, in this work we focus on advantages for the most difficult type of information transmission.

Comparing the superposition schemes from Fig.~\ref{cartoon} in the case of a classical communication yields an analogous behavior to the case of quantum communication. It was already shown in Appendix G of Ref.~\cite{gisin2005}, that superpositions of quantum channels [Fig.~\ref{cartoon}\textbf{a)}] yield at least the amount of transmissible classical information of classical trajectories [Fig.~\ref{cartoon}\textbf{d)}]. The same holds true for the advantages through the quantum-control of channel order [Fig.~\ref{cartoon}\textbf{c)}] considered in the initial work~\cite{Ebler2018}. Later, Ref.~\cite{abbott2018} showed that quantum-control of parallel channels [Fig.~\ref{cartoon}\textbf{a)}] can outperform the quantum-control of channel order [Fig.~\ref{cartoon}\textbf{c)}] for certain communication tasks. Numerical simulations support generality of this claim for classical information by comparing the two scenarios for random channels~\cite{Grinbaum2020}. 
Finally, quantum controlled sequences of channels [Fig.~\ref{cartoon}\textbf{b)}] were found to allow for the highest classical communication rates, as it has a larger set of allowed encoding schemes, which allow for phase kick-backs that cause partial information exchange with the trajectory degree of freedom~\cite{guerin2019}.

\subsection{Case Study: Activation of the EB-Channels in the Three Layouts}
\label{subsec:calc}

In this section, we briefly evaluate the output state of the three superposition techniques for the noisy channel described by Eq.~\eqref{eq:XYsimp}.
We carry out our study by interpreting the action of the two channels as follows.
Each channel randomly applies either $X$ or $Y$ with probability $1/2$. 
Hence, the `internal configuration' of the superposition can be understood as either \textit{i.}\ both channels 1 and 2 implement $X$, \textit{ii.}\ channel 1 implements $X$, while channel 2 implements $Y$, \textit{iii.}\ channel 1 implements $Y$, whereas channel 2 implements $X$, or \textit{iv.}\ both channels 1 and 2 implement $Y$.
The final output state will then be a mixture of the output states in these four configurations, each with probability $1/4$. 

Throughout this section, we will assume that the input state is $\ket{\psi}_\mathrm{I}\ket{+}_\mathrm{T}$, where $\ket{\psi}=\alpha\ket{0} +\beta\ket{1}$, and $I$ ($T$) labels the internal (trajectory) DOF.
 
\vspace{3mm} 
 
\emph{Quantum-Control of Parallel Channels---}  In this scheme, one noisy channel is placed in each trajectory.
The action of the two channels in parallel can be interpreted as creating a mixture of the following four unnormalised states at the output:
\begin{subequations}
\begin{align}
X\ket{\psi}_\mathrm{I}\ket{0}_\mathrm{T} &+ X\ket{\psi}_\mathrm{I}\ket{1}_\mathrm{T}\\
X\ket{\psi}_\mathrm{I}\ket{0}_\mathrm{T} &+ Y\ket{\psi}_\mathrm{I}\ket{1}_\mathrm{T}\\
Y\ket{\psi}_\mathrm{I}\ket{0}_\mathrm{T} &+ X\ket{\psi}_\mathrm{I}\ket{1}_\mathrm{T}\\
Y\ket{\psi}_\mathrm{I}\ket{0}_\mathrm{T} &+ Y\ket{\psi}_\mathrm{I}\ket{1}_\mathrm{T},
\end{align}
\end{subequations}
where the four states correspond to internal configurations of $X-X$, $X-Y$, $Y-X$, and $Y-Y$, for channel 1 and channel 2, respectively.
These states can be rewritten (up to further normalisation) as:
\begin{subequations}
\begin{align}
X\ket{\psi}_\mathrm{I}&\ket{+}_\mathrm{T}\\
(X\ket{\psi}_\mathrm{I}+Y\ket{\psi}_\mathrm{I})\ket{+}_\mathrm{T} &+ (X\ket{\psi}_\mathrm{I}-Y\ket{\psi}_\mathrm{I})\ket{-}_\mathrm{T}\\
(X\ket{\psi}_\mathrm{I}+Y\ket{\psi}_\mathrm{I})\ket{+}_\mathrm{T} &- (X\ket{\psi}_\mathrm{I}-Y\ket{\psi}_\mathrm{I})\ket{-}_\mathrm{T}\\
Y\ket{\psi}_\mathrm{I}&\ket{+}_\mathrm{T}.
\end{align}
\end{subequations}
Now, measuring the trajectory DOF in the $\{\ket{+}_\mathrm{T},\ket{-}_\mathrm{T}\}$ basis and obtaining $\ket{-}_\mathrm{T}$ (which happens with probability 1/4) projects the internal DOF into $X\ket{\psi}_\mathrm{I}-Y\ket{\psi}_\mathrm{I}$, which can be rewritten (after renormalisation and up to an irrelevant global phase) as in Eq.~\eqref{eq:minusPS}.
This is a pure state, which implies that some ability to transmit quantum information has been restored in post-selection.
If, on the other hand, one obtains the result $\ket{+}_\mathrm{T}$ (with probability 3/4), it is straightforward to show that the internal DOF is projected in the mixed state described by Eq.~\eqref{eq:plusPS}.

In the recent papers on superpositions of trajectories~\cite{abbott2018,chiribella2019quantum}, it was shown that the output of a  quantum-controlled superposition of two channels depends on additional parameters related to the physical realisation of the channels (`transformation matrices' in~\cite{abbott2018} and `vacuum amplitudes' in~\cite{chiribella2019quantum}). In our scheme, these additional parameters reduce to the relative phase between the vacuum and the single-photon subspace of the unitary operations (\textit{e.g.}, the Pauli-$X$ and -$Y$ from above, with transformation matrix $\Gamma = (X+Y)/2$). More precisely, the vacuum extension of a qubit unitary $U$ is $U' = e^{i \phi} \ket{\text{vacuum}}\bra{\text{vacuum}} + U$, where $U$ acts in the single-photon subspace. In the calculation above, the phase is implicitly set to zero, which is in agreement with our experiment.

\vspace{3mm} 

\emph{Channels in Series with Quantum-Controlled Operations---}
Let us now consider the action of the superposition of trajectories in series with the quantum-controlled operations (Fig.~\ref{cartoon}\textbf{b)}), with $U_1=Y$, $U_2=\mathcal{I}$, $U_3=\mathcal{I}$.
In this case, the input state is transformed into $(\ket{\psi}_\mathrm{I}\ket{0}_\mathrm{T} + Y\ket{\psi}_\mathrm{I}\ket{1}_\mathrm{T})/\sqrt{2}$ before interacting with the noisy channels.

Again, we can compute the four effective unnormalised states which arise from the different internal configurations of the noisy channels:
\begin{subequations}
\begin{align}
XX\ket{\psi}_\mathrm{I}\ket{0}_\mathrm{T} &+ XXY\ket{\psi}_\mathrm{I}\ket{1}_\mathrm{T}\\
YX\ket{\psi}_\mathrm{I}\ket{0}_\mathrm{T} &+ YXY\ket{\psi}_\mathrm{I}\ket{1}_\mathrm{T}\\
XY\ket{\psi}_\mathrm{I}\ket{0}_\mathrm{T} &+ XYY\ket{\psi}_\mathrm{I}\ket{1}_\mathrm{T}\\
YY\ket{\psi}_\mathrm{I}\ket{0}_\mathrm{T} &+ YYY\ket{\psi}_\mathrm{I}\ket{1}_\mathrm{T}.
\end{align}
\end{subequations}
The order of the above states refers to internal configurations $X-X$, $X-Y$, $Y-X$, and $Y-Y$.
These states can be rewritten (up to phases) as:
\begin{subequations}
\begin{align}
\ket{0}_\mathrm{I}\ket{\psi'}_\mathrm{T} &+i \ket{1}_\mathrm{I} X\ket{\psi'}_\mathrm{T}\\
\ket{0}_\mathrm{I}\ket{\psi'}_\mathrm{T} &-i \ket{1}_\mathrm{I} X\ket{\psi'}_\mathrm{T}\\
\ket{0}_\mathrm{I}\ket{\psi'}_\mathrm{T} &-i \ket{1}_\mathrm{I} X\ket{\psi'}_\mathrm{T}\\
\ket{0}_\mathrm{I}\ket{\psi'}_\mathrm{T} &+i \ket{1}_\mathrm{I} X\ket{\psi'}_\mathrm{T},
\end{align}
\end{subequations}
where $\ket{\psi'}=\alpha\ket{0}-i\beta\ket{1}$.
As a result, we see that measuring the internal DOF in the computational basis $\{\ket{0}_\mathrm{I},\ket{1}_\mathrm{I}\}$ projects the trajectory into either $\ket{\psi'}_\mathrm{T}$ or $X\ket{\psi'}_\mathrm{T}$, upon obtaining outcomes $\ket{0}_\mathrm{I}$ or $\ket{1}_\mathrm{I}$, respectively (each with equal probabilities).
Both of these states can be unitarily corrected, allowing one to achieve perfect quantum information transfer through these channels. (Even though here, for simplicity, we restricted ourselves to the case $p=1/2$, the same reasoning applies whatever the mixing probability $p$ in the definition of the channels, Eq.~\eqref{XY}).

\vspace{3mm} 

\emph{Quantum-Control of Channel Order---}
For this scheme, we make use of the fact that the output of the quantum switch for unitary operations $A$ and $B$ is
\begin{equation}
\frac{1}{2}\{A,B\}\ket{\psi}_\mathrm{I}\ket{+}_\mathrm{T} - \frac{1}{2}[A,B]\ket{\psi}_\mathrm{I}\ket{-}_\mathrm{T},
\end{equation}
where $[A,B]$ is the commutator of $A$ and $B$, and $\{A,B\}$ is their anti-commutator.
It is then easy to notice that the four output states (up to phases) are
\begin{subequations}
\begin{align}
\ket{\psi}_\mathrm{I}&\ket{+}_\mathrm{T}\\
Z\ket{\psi}_\mathrm{I}&\ket{-}_\mathrm{T}\\
Z\ket{\psi}_\mathrm{I}&\ket{-}_\mathrm{T}\\
\ket{\psi}_\mathrm{I}&\ket{+}_\mathrm{T}.
\end{align}
\end{subequations}
Again, the order of the above states refers to the internal configurations $X-X$, $X-Y$, $Y-X$, and $Y-Y$.
This leads to the mixture described by Eq.~\eqref{eq:outputSWITCH}, and it implies that measuring the trajectory in the $\{\ket{+}_\mathrm{T},\ket{-}_\mathrm{T}\}$ basis projects the internal DOF into either $\ket{\psi}_\mathrm{I}$, or $Z\ket{\psi}_\mathrm{I}$. (As above, the same reasoning also applies whatever the mixing probability $p$ in Eq.~\eqref{XY}.)

\subsection{Liquid Crystals Characterization}
\label{subsec:LiquidCrystals}

\begin{figure}
  \centering%
      \includegraphics[width=\columnwidth]{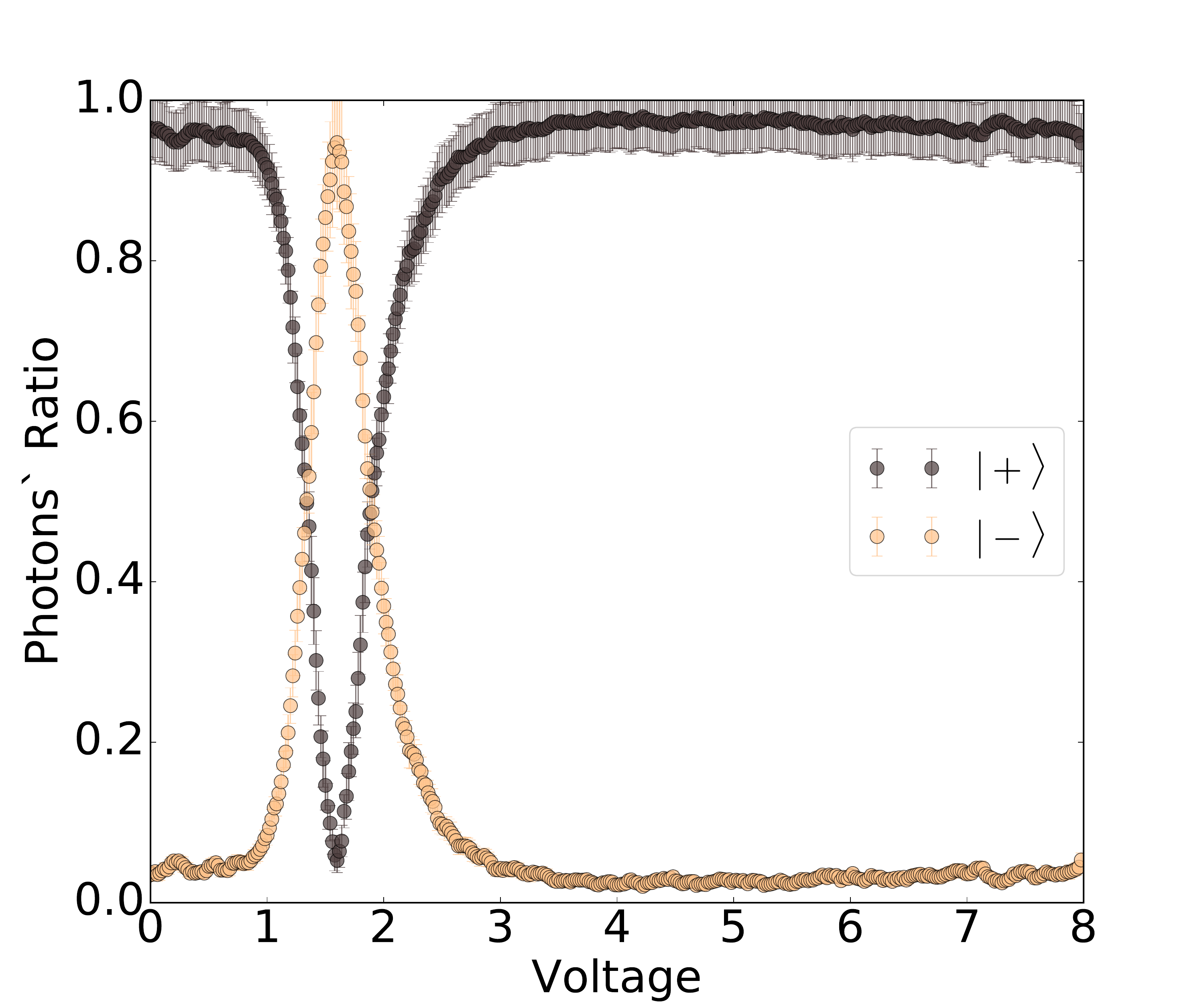}
  \caption{\textbf{Experimental characterization of a liquid crystal waveplate (LCWP) at $0^\circ$.} Since the crystal is positioned at $0^\circ$, it will be able to switch from an identity operation to a Pauli-$Z$. To characterize the voltage corresponding to a Pauli-$Z$, we send through it photons in the polarization basis $\{\vert \pm \rangle = (\vert 0 \rangle \pm \vert 1 \rangle)/\sqrt{2}\}$, and we measure for which voltage the population inversion occurs. The estimated errors are Poissonian.}
    \label{fig:LC_Charact}
\end{figure}

In essence, a liquid crystal waveplate (LCWP) can be understood as a standard crystalline retarder whose amount of retardance can be continuously varied by applying a voltage. Fig.~\ref{fig:LC_Charact} shows the characterization of one of our LCWPs. In our experiment, we used Meadowlark Liquid Crystal Variable Retarders. Their beam deviation is estimated to 2 arc min, their reflectance (per surface) is 0.5\%, and their surface quality is 40-20 scratch and dig.

\subsection{Numerical Comparison for Random Channels}
\label{subsec:Random_Channels}

\begin{figure}[bt]
  \centering%
      \includegraphics[width=\columnwidth]{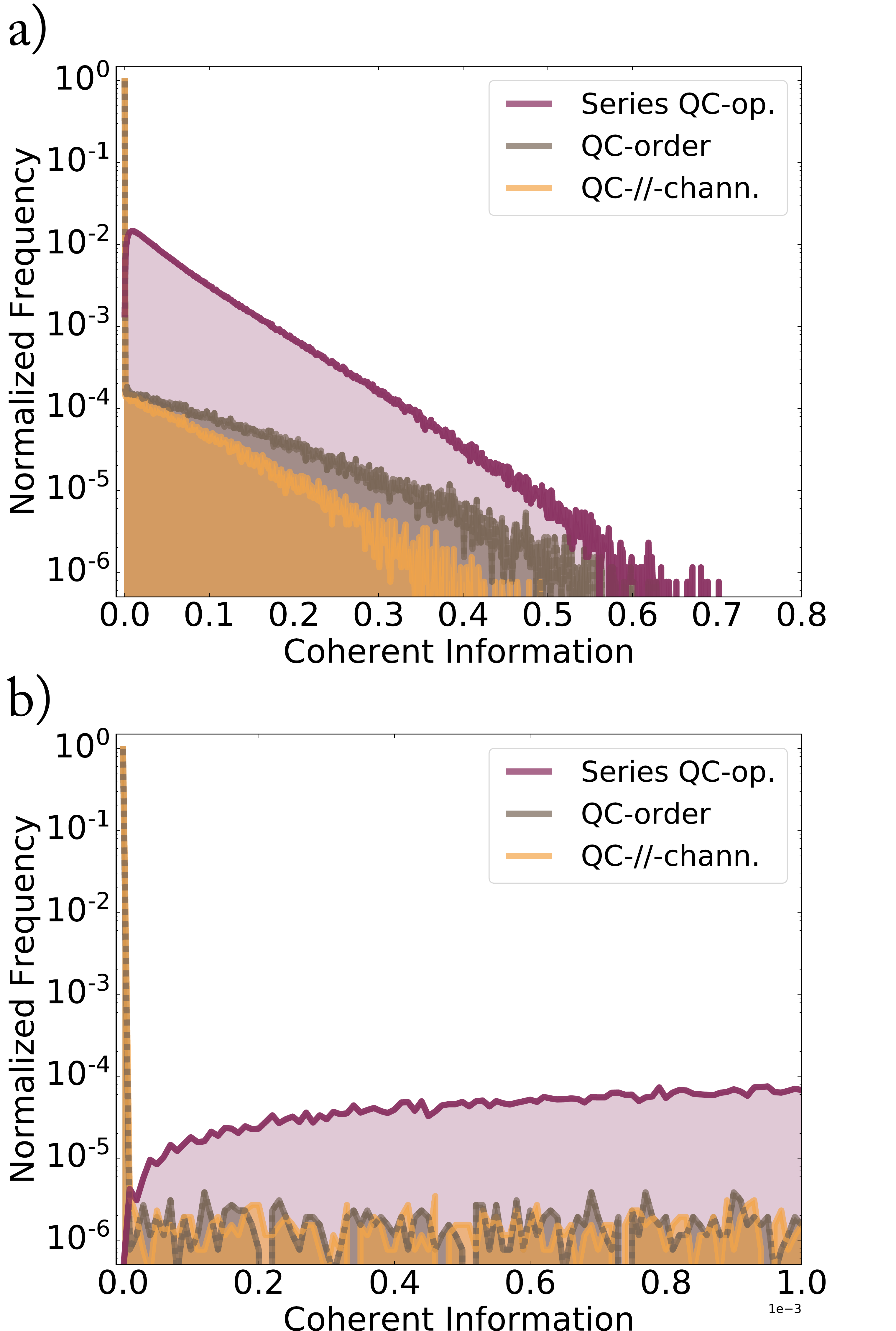}
  \caption{\textbf{Histogram of the coherent information achieved with the three channel layouts when the two copies of the same randomly-generated channel is used in each of the three layouts.} The histograms report the frequency with which a random channel (\textit{y}-axis, in logarithmic scale) yields a given amount of coherent information (\textit{x}-axis), normalized to the total number of channels used. \textbf{a) Histogram with $10^3$ bins between a coherent information of 0 and 0.85.} 
As can be seen, the configuration of channels in series with quantum-controlled operations consistently achieves the highest coherent information on average. 
\textbf{b)  Histogram of the same data with $10^5$ bins displayed for values of coherent information from 0 to 0.001.} By increasing the resolution for small values of coherent information, it is possible to observe in greater detail the absence of the peak at zero for the quantum superposition of channels in series with quantum-controlled operations. In this region, the performance of the quantum-control of parallel channels and that of quantum-control of channel order is comparable.}
    \label{fig:Random_Channels}
\end{figure}

\begin{figure}
  \centering%
      \includegraphics[width=\columnwidth]{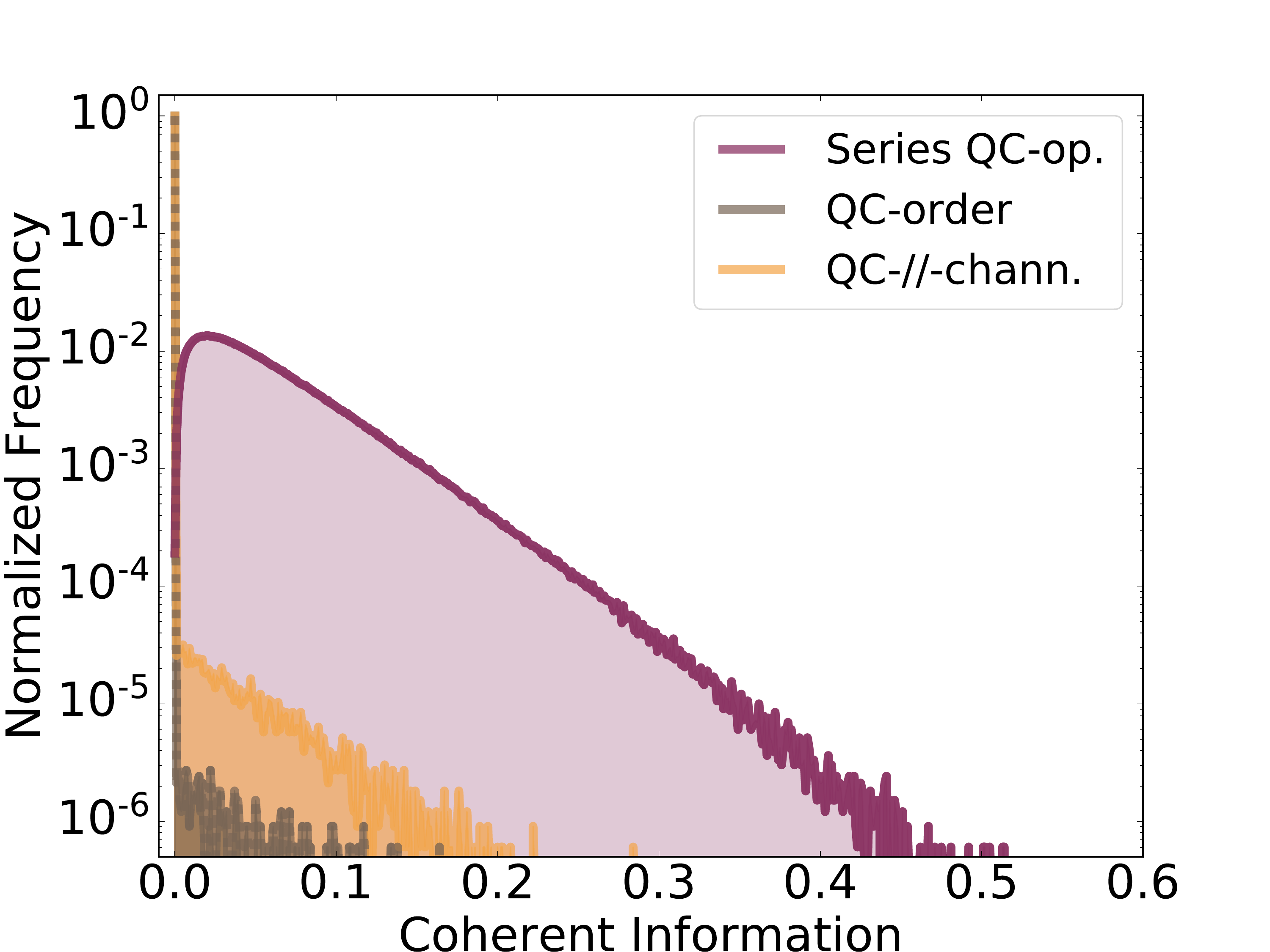}
  \caption{\textbf{Histogram of the coherent information achieved with the three channel layouts when two different randomly-generated channels are used.} The overall trend here is comparable to that of two copies of the same random channel (Fig. \ref{fig:Random_Channels}).
  However, in this case, the quantum-control of the parallel channels performs, on average, better than the  quantum-control of channel order. Moreover, in general, all three layouts tend to perform worse than in the case of two copies of the same random noisy channel (\textit{i.e.}, the maximum amount of coherent information which can be achieved through each layout is generally lower than in the case shown in Fig.~\ref{fig:Random_Channels}).}
    \label{fig:Two_Random_Channels}
\end{figure}

To further compare the three schemes, we present a numerical evaluation of the coherent information which can be achieved with each channel layout for a large set of randomly-generated channels. The numerical procedure is carried out as follows. First, we randomly generate a quantum completely-positive and trace-preserving (CPTP) channel using the \texttt{quantinf MATLAB} package available at~\cite{quantinf_matlab}. (The package uses the routine outlined in~\footnote{We made use of the \texttt{randChan}
 function of the \texttt{quantinf MATLAB} package, which returns a randomly generated $2\times 2$ quantum channel in the Kraus representation. It achieves this by generating a random isometry, and converting it onto the Kraus representation. Below, we provide further details on how these two tasks are achieved. \newline \textbf{\textit{Step 1)}} The program generates a random isometry in the following way: 
\textit{Step 1.1)} it generates two $2\times 2$ matrices, $A$ and $B$, of real floating-point random entries drawn from a standard normal distribution (\textit{i.e.}, $\mu = 0$, $\sigma = 1$, being $\mu$ the mean of the distribution, and $\sigma$ its variance) using the \texttt{randn} function; \textit{Step 1.2)} it builds $C = (A+iB)/\sqrt{2}$; \textit{Step 1.3)} it performs an economy-size QR decomposition such that $C = Q*R$ (where $Q$ is an orthogonal matrix, and $R$ is an upper triangular matrix); \textit{Step 1.4)} it diagonalises and normalises the matrix $R$ into $R'=\text{diag}\bigl(\text{diag}(R)/\vert\text{diag}(R)\vert\bigr)$; \textit{Step 1.5 } it creates a new matrix $V = Q*R'$, which is the desired isometry. \newline \textbf{\textit{Step 2)}} The program converts the isometry $V$ into the Kraus representation by left-multiplying it by states in the computational basis.}.) Then, we estimate the coherent information when \textit{i.} two copies of the same channel (Fig.~\ref{fig:Random_Channels})  or \textit{ii.} two different randomly-generated channels (Fig.~\ref{fig:Two_Random_Channels}) are inserted in the three types of quantum superpositions of trajectories.
The coherent information of the resulting superposition is then estimated using a maximally entangled state $\vert\Phi^{+}\rangle$ as input~\footnote{As before, the choice of a maximally-entangled state lower bounds the quantum capacity, and numerical simulations suggest that such states maximise the amount of coherent information of a given channel.}. For the configuration of quantum-control of parallel channels, as shown in~\cite{abbott2018}, the output state depends not only on the CPTP map, but also on the specific implementation of the channel. We therefore discuss three methods to generate different implementations for each randomly generated CPTP map, but we present results only from the one corresponding to the experimental implementation reported in the main text. 
For the configuration of channels in series with quantum-controlled operations, we set $U_1=Y$, and $U_2 = U_3 = \mathcal{I}$. Further details on our numerical procedure are reported in Subsection~\ref{subsubsec:Sum_Numerics}.
Histograms of our results with respect to the coherent information are presented in Figs.~\ref{fig:Random_Channels}-\ref{fig:Two_Random_Channels}, for $2.6 \times 10^6$ iterations (Fig.~\ref{fig:Random_Channels}) and $3.3 \times 10^6$ iterations (Fig.~\ref{fig:Two_Random_Channels}). 
In essence, this can be interpreted as the probability to obtain a given value of coherent information with each of the three layouts.  We observe that, on average, channels in series with quantum-controlled operations achieve a better performance than the other two methods, while the quantum-control of parallel channels (quantum-control of channel order) exhibits the lowest performance when operated with two copies of the same channel (two different randomly-generated channels). 
We also note that, in many cases, both the quantum-control of parallel channels and the quantum-control of channel order fail to obtain any activation, leading to large peaks at zero in each of their histograms.
Interestingly, this peak is not present in the case of the layout in series with quantum-controlled operations. This suggests that an activation (albeit small) of the noisy channels can always be achieved using this layout with only two trajectories.

\begin{figure}[htb]
  \centering%
      \includegraphics[width=\columnwidth]{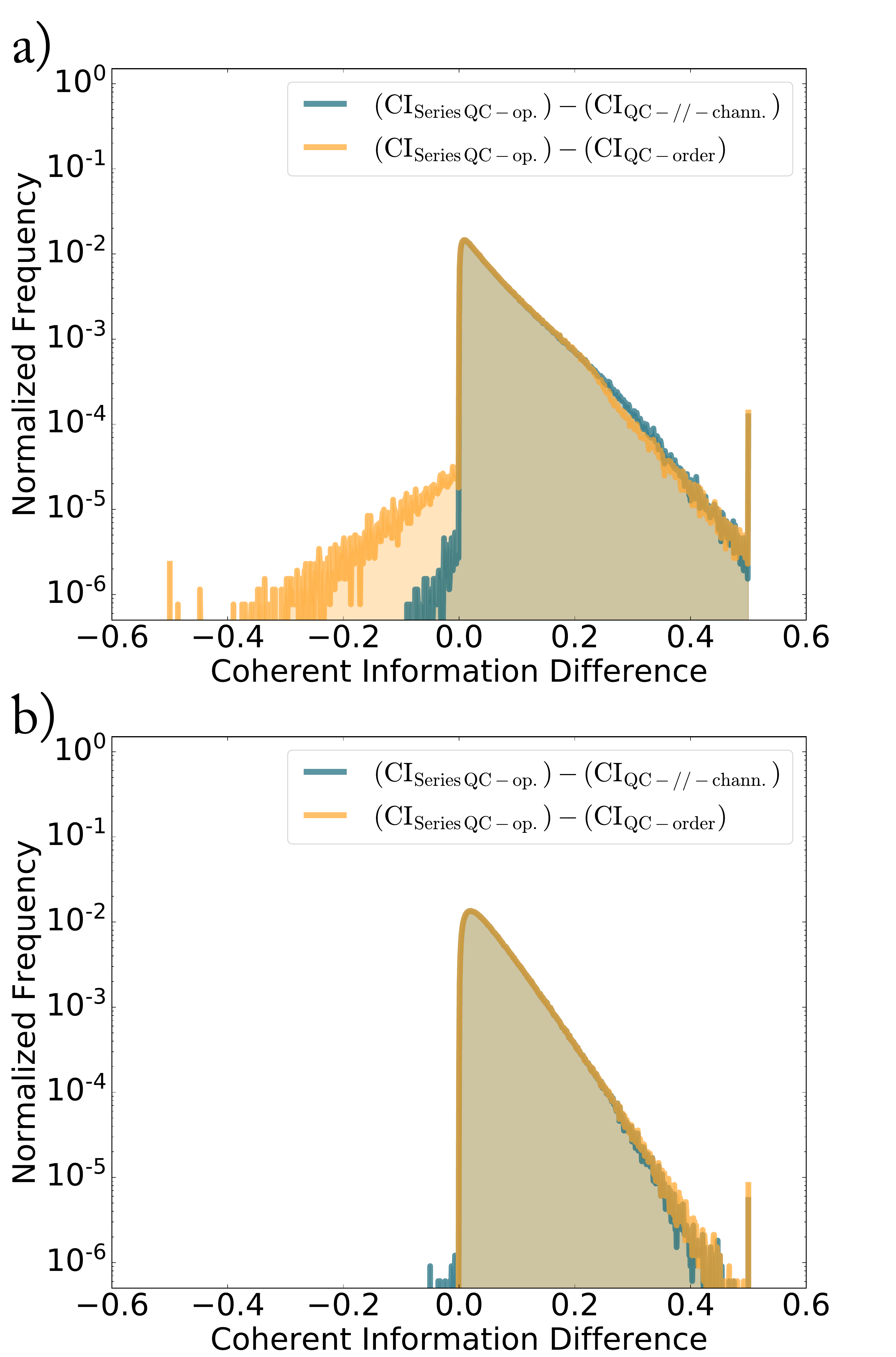}
  \caption{\textbf{Histograms of the difference between coherent information achievable with quantum superposition of channels in series with quantum-controlled operations and the other two layouts in the case of a) two independent copies of the same random channel, and b) two different randomly-generated channels.} The histograms show, for each random channel, the difference between the coherent information of the quantum superposition of channels in series with quantum-controlled operations and that of the quantum-control of parallel channels [$(\text{CI}_{\text{Series QC-op.}})-(\text{CI}_{\text{QC-//-chann.}})$] and of quantum-control of channel order [$(\text{CI}_{\text{Series QC-op.}})-(\text{CI}_{\text{QC-order}})$]. While, to a large extent, the layout using the channels in series with quantum-controlled operations tends to outperform the other two schemes, the negative values indicate that this is not always the case.}
    \label{fig:Random_Channels_Difference}
\end{figure}

Finally, Fig.~\ref{fig:Random_Channels_Difference} shows a histogram, wherein the difference between the coherent information of the quantum superposition of channels in series with quantum-controlled operations and that of the quantum-control of parallel channels ($\text{CI}_{\text{Series QC-op.}}-\text{CI}_{\text{QC-//-chann.}}$) and of quantum-control of channel order ($\text{CI}_{\text{Series QC-op.}}-\text{CI}_{\text{QC-order}}$) is plotted for each random pair of channels. Generally, the quantum superposition of channels in series with quantum-controlled operations can achieve coherent information values higher than two other layouts. However, the negative values in the histograms show that this is not always the case. This is in line with what illustrated in Section~\ref{subsec:Fixing_Operations}, where we highlight the fact that, if the unitaries $U_1$, $U_2$ and $U_3$ were not optimised for given noisy channels, a higher coherent information might be obtainable with the other layouts. While Ref.~\cite{guerin2019} proved that, by superposing a larger number of trajectories, one can always find an optimal choice of quantum-controlled unitaries which can outperform the other two channels' layouts, we leave it as an open question whether or not this is also true for the restricted case of two trajectories only.

\subsubsection{Summary of Numerics}
\label{subsubsec:Sum_Numerics}

In this subsection, we provide further details on how we constructed the output states in the three schemes. We start by randomly generating two single-qubit channels, using the Kraus decomposition. Since any qubit channel has a Kraus decomposition with 4 operators or less, this results in two sets of Kraus operators $\{A_0, A_1, A_2, A_3\}$ and $\{B_0, B_1, B_2, B_3\}$.
We will always use the two-qubit Bell state $\ket{\Phi^+}$ to probe the channel and calculate the coherent information.
Then, the full three-qubit state we consider is given by:
\begin{equation}
\ket{\psi^\mathrm{in}}_{\mathrm{T},\mathrm{I},\mathrm{H}}=\ket{+}_{\mathrm{T}}\otimes\ket{\Phi^+}_{\mathrm{I},\mathrm{H}},
\end{equation}
where $\mathrm{T}$ is the trajectory qubit, $\mathrm{I}$ is the system which will experience the noisy channel (information qubit), and $\mathrm{H}$ is the auxiliary (hypothetical) qubit used to evaluate the coherent information. We will use $A_i^{(\mathrm{I})}=A_i\otimes\mathcal{I}$ as shorthand, where $A_i$ acts on the state of the information qubit and $\mathcal{I}$ on that of the auxiliary qubit.

\emph{Quantum-Control of Parallel Channels---}
We construct the output state, following Ref.~\cite{abbott2018}, as
\begin{align}
\rho^\mathrm{out}_{\mathrm{T},\mathrm{I},\mathrm{H}}=&\frac{1}{2}\bigl[ \ket{0}\bra{0}_\mathrm{T}\otimes \mathcal{C}_A(\rho^\mathrm{in}_{\mathrm{I},\mathrm{H}}) + \ket{1}\bra{1}_\mathrm{T}\otimes \mathcal{C}_B(\rho^\mathrm{in}_{\mathrm{I},\mathrm{H}}) \bigr] \\
+& \frac{1}{2}\left[\ket{0}\bra{1}_\mathrm{T}\otimes \Gamma_A\,\rho^\mathrm{in}_{\mathrm{I},\mathrm{H}}\, \Gamma_B^\dagger +\ket{1}\bra{0}_\mathrm{T}\otimes \Gamma_B\;\rho^\mathrm{in}_{\mathrm{I},\mathrm{H}}\, \Gamma_A^\dagger \right],\notag
\end{align}
where $\rho^\mathrm{in}_{\mathrm{I},\mathrm{H}} = \ket{\Phi^+}\bra{\Phi^+}_{\mathrm{I},\mathrm{H}}$. Here, $\mathcal{C}_A$ and $\mathcal{C}_B$ are the application of either channel
\begin{equation}
\mathcal{C}_\Xi(\rho^\mathrm{in}_{\mathrm{I},\mathrm{H}})=\sum_{i=0}^3 \Xi_i\,\rho^\mathrm{in}_{\mathrm{I},\mathrm{H}}\,\Xi_i^\dagger
\end{equation}
with $\Xi_i = A_i, B_i$, and the transformation matrices $\Gamma_\Xi$ are
\begin{equation}
\Gamma_\Xi=\sum_{i=0}^3 \braket{E_\Xi \vert i}\Xi_i,
\end{equation}
where $\{\ket{i}\}$ are orthogonal states of the environment. The transformation matrices $\Gamma_\Xi$ are related to a specific purification of the channels, and they depend on the initial states of the environment $\ket{E_A}$ and $\ket{E_B}$ used in this purification.  
These states will be given by the actual physical implementation of the channel, and they can lead to different activations using the quantum-controlled channels.
Given some Kraus representations of the channels, we numerically investigated three different states of the environment.
First, as used in Ref.~\cite{abbott2018}, we set $\ket{E_A}=\ket{E_B}=\frac{1}{2}\sum_{i=0}^3\ket{i}$.
Second, we generate $\ket{E_A}$ and $\ket{E_B}$ randomly from the Haar measure for each different channel.
In this case we do not optimise over $\ket{E_A}$ and $\ket{E_B}$, we simply take one random state for each.
Finally, we set the weights of the environment based on the randomly-generated channel as:
\begin{equation}
\ket{E_\Xi}=\sum_{i=0}^3\sqrt{w^{(\Xi)}_i}\ket{i},
\end{equation}
where
\begin{equation}
w^{(\Xi)}_i=\mathrm{Tr}\Bigl(\Xi_i^{(\mathrm{I})}\rho^\mathrm{in}_{\mathrm{I},\mathrm{H}}{\Xi_i^{(\mathrm{I})}}^\dagger\Bigr),
\end{equation}
which, in our case of a maximally entangled input state, reduces to $\mathrm{Tr}\bigl(\Xi_i^{(\mathrm{I})}{\Xi_i^{(\mathrm{I})}}^\dagger\bigr)/2$.

The appropriate method to generate the states of the environment depends on the physical realisation of the channels. 
Within our framework, the description of quantum-control of parallel channels given in the main text, for the channels we realised experimentally, coincides with the third option, with $\Xi_i=w_i \sigma_i$, where $\sigma_i$ is a Pauli unitary, and the weights $w_i$ are given by the coefficients in Eqs.~\eqref{XY}--\eqref{BB84}.
Correspondingly, we present the results for this case in Figs.~\ref{fig:Random_Channels}--\ref{fig:Random_Channels_Difference}~\footnote{Although an optimisation over all environmental states may produce higher values of the coherent information, the choice of environmental states presented in Figs.~\ref{fig:Random_Channels}--\ref{fig:Random_Channels_Difference} corresponds to the one which was experimentally realised with our photonic Mach-Zehnder interferometer without further control over the environment or the noisy channels. To adhere to the spirit of experimentally comparing schemes pursued in this paper, we thus chose not to optimise explicitly over all implementations of the randomly chosen channels (following the approach, \textit{e.g.}, of Ref.~\cite{abbott2018}).}. Moreover, on average the final method (\textit{i.e.}, setting the state of the environment based on the Kraus operators) performs the best among the aforementioned three. This can perhaps be explained by the fact that, compared to the other ones, this method generally leads to a larger norm of the transformation matrices $\Gamma_\Xi$, which is crucial for the communication advantages~\cite{abbott2018, kristjansson2019, kristjansson2020single}.

\emph{Channels in Series with Quantum-Controlled Operations---} 
For the numerical evaluation of this layout, we must also include in the description the controlled-unitaries which are applied before the noisy channels:
\begin{equation}
\text{C-U}=\ket{0}\bra{0}_\mathrm{T}\otimes\mathcal{I} + \ket{1}\bra{1}_\mathrm{T}\otimes{U}. 
\end{equation}
For all of the numerical results presented here, we set $U=Y$, where $Y$ is the Pauli-\textit{Y} operator.
Then, we construct 16 combined Kraus operators
\begin{equation}
K_{i,j}=B_j^{(\mathrm{I})}A_i^{(\mathrm{I})},
\end{equation}
and when, \textit{e.g.}, the initial trajectory state is $\ket{+}_\mathrm{T}$, we compute the output state as
\begin{equation}
\rho^\mathrm{out}_{\mathrm{T},\mathrm{I},\mathrm{H}}= \frac{1}{2} \sum_{k,l} U^k \ket{k}\bra{l}_\mathrm{T} (U^\dagger)^l \otimes \sum_{i,j} K_{i,j} \rho^\mathrm{in}_{\mathrm{I},\mathrm{H}} K_{i,j}^\dagger,
\end{equation}
where we used the notation according to which $U^0 = \mathcal{I}$.

\emph{Quantum-Control of Channel Order---} 
For the switch we will use a simplification. We know that for a given pair of Kraus operators, the output state when the trajectory is prepared in $\ket{+}_\mathrm{T}$ and the input is one half of the maximally entangled state $\ket{\Phi^+}_{\mathrm{I},\mathrm{H}}$ is:
\begin{align}
\rho_{i,j}=&\bigl\{A_i^{(\mathrm{I})},B_j^{(\mathrm{I})}\bigr\}\,\rho^+_{\mathrm{T},\mathrm{I},\mathrm{H}}\,\bigl\{A_i^{(\mathrm{I})},B_j^{(\mathrm{I})}\bigr\}^\dagger \notag\\
+& \bigl[A_i^{(\mathrm{I})},B_j^{(\mathrm{I})}\bigr]\,\rho^-_{\mathrm{T},\mathrm{I},\mathrm{H}}\,\bigl[A_i^{(\mathrm{I})},B_j^{(\mathrm{I})}\bigr]^\dagger,
\end{align}
where $\{A_i^{(\mathrm{I})},B_j^{(\mathrm{I})}\}$ and $[A_i^{(\mathrm{I})},B_j^{(\mathrm{I})}]$ are the anti-commutator and commutator of $A_i^{(\mathrm{I})}$ and $B_j^{(\mathrm{I})}$, and $\rho^+_{\mathrm{T},\mathrm{I},\mathrm{H}}=\ket{+}\bra{+}_\mathrm{T}\otimes\ket{\Phi^+}\bra{\Phi^+}_{\mathrm{I},\mathrm{H}}$ and $\rho^-_{\mathrm{T},\mathrm{I},\mathrm{H}}=\ket{-}\bra{-}_\mathrm{T}\otimes\ket{\Phi^+}\bra{\Phi^+}_{\mathrm{I},\mathrm{H}}$.
Then the net output state is simply
\begin{equation}
\rho^\mathrm{out}_{\mathrm{T},\mathrm{I},\mathrm{H}}=\sum_{i,j}\rho_{i,j}.
\end{equation}
From these output states we then evaluate the coherent information as described in the main text.


\subsection{Data Acquisition and Error Estimation}
\label{subsec:Data_Acquisition}

Below, we briefly outline the details of the data acquisition and the error estimation in our experiment.

As discussed above, we followed two methods to experimentally construct of the noisy channels. 
In the first, we realized the noise within each channel by generating random numbers in the range $[0,1]$.
Based on this number, and on the type of noise we wanted to realize (Eqs.~\eqref{XY}--\eqref{BB84}), we assigned a unitary operation from the set $\{\mathcal{I}, X, Y, Z\}$. 
In order to ensure a high fidelity of the noise channel ($> 99\%$), we repeated this procedure 1000 times, measuring each configuration for 1s, and integrating the data taking procedure over these 1000 runs. 
In the second method, we measured all the possible combinations of unitary operations between the two noisy channels 1 and 2, and we then created the desired noise during our data analysis, following the procedure proposed in Ref.~\cite{goswami2018communicating}.
The first method was used to create the noisy channels in the indefinite order channel layout, while the second method was used for all other layouts. 
We did not observe any significant difference in the performance of the two methods (provided that we applied enough random unitary operations, see Section \ref{subsec:Process_Fidelity}).
However, the first method required several days of measurement, the second less than an hour.
Since full quantum process tomography (QPT) was not required for the indefinite order arrangement, we only used the first method for these data, and used the second method for all of the remaining channel configurations. 

We collect $\approx$ 23 000 entangled photon pairs per second directly from our source. Of these pairs, we selected only one separable polarization component (\textit{i.e.}, $\ket{H,V}$), halving the count rate.
Finally, the photons were sent through optical fibers to the different experiments. 
Because of experimental imperfections due to the non-zero reflectivity of the various optical elements, the non-ideal fiber coupling, and the optical fiber's losses (the distance to travel in optical fiber between the source and the various experiments is about 3m), approximately 3000 photons per second were detected at the end of the experiment. 

Finally, because of the long measurement times (particularly, in the case of the physical implementation of the noise in the channels), we observed phase drifts in the two arms of the interferometer. In order to correct these drifts, so as to ensure that we always prepared and projected the desired path qubit states, we actively stabilized the interferometer by means of the delay line controlled by a piezo-actuator. We measured and reset the phase every 20 minutes (which, according to our tests, ensured phase drifts below 1\%). Given these count rates, it follows that we measured about $3 \times 10^6$ photons for each internal configuration in case of physical implementation of the noisy channels, and about 3000 counts for each internal configuration in case of implementation of the noisy channels during data analysis.

The aforementioned imperfections in the path  qubit had various consequences in our experiment. 
First, the phase drift on this qubit caused an uncertainty on the input state, since the phase of the path qubit can fluctuate over time.
Moreover, if the phase drifts during the experiment, the purity of the input state can be reduced.
In light of this, and of the high number of accumulated counts, the main statistical error in our experiment was related to the input state used for QPT.
Therefore, to calculate all our experimental error bars, we determined the input state as follows.
We performed quantum tomography of the quantum state directly on the path and polarization qubits in absence of `internal operations' (\textit{i.e.}, setting to identity all the optical components meant to implement the noisy channels later on).
We then observed the variation of this state as a function of time, and used this variation to analyze our data for a `worst-case' state, a `best-case' state, and a `most-likely' state.
This spread in the input states led to the error bars and the data-points presented in all of our experimental data.

The second main cause of experimental errors in the path qubit arises from the imperfect visibility of the Mach-Zehnder interferometers used to measure it.
This reduced visibility, $\approx$ 0.93--0.96 (depending on the amount of optical elements in each interferometer arm) essentially corresponds to a slight loss of coherence in the two trajectories and, therefore, to a decrease in the ability of the trajectories to restore quantum information.
This leads to a systematic offset in our data.
Likewise, a slight miscalibration of our waveplates and liquid crystal waveplates may have occurred, this would lead to further systematic errors.
These systematic errors have not been included in our error bar calculation.
We note, in particular, that a considerably reduced visibility of the interferometer may constitute the greatest experimental challenge in the application of our techniques to real-world quantum communication.

\subsection{Fidelity of Channel Implementation}
\label{subsec:Process_Fidelity}

\begin{figure}
  \centering%
      \includegraphics[width=\columnwidth]{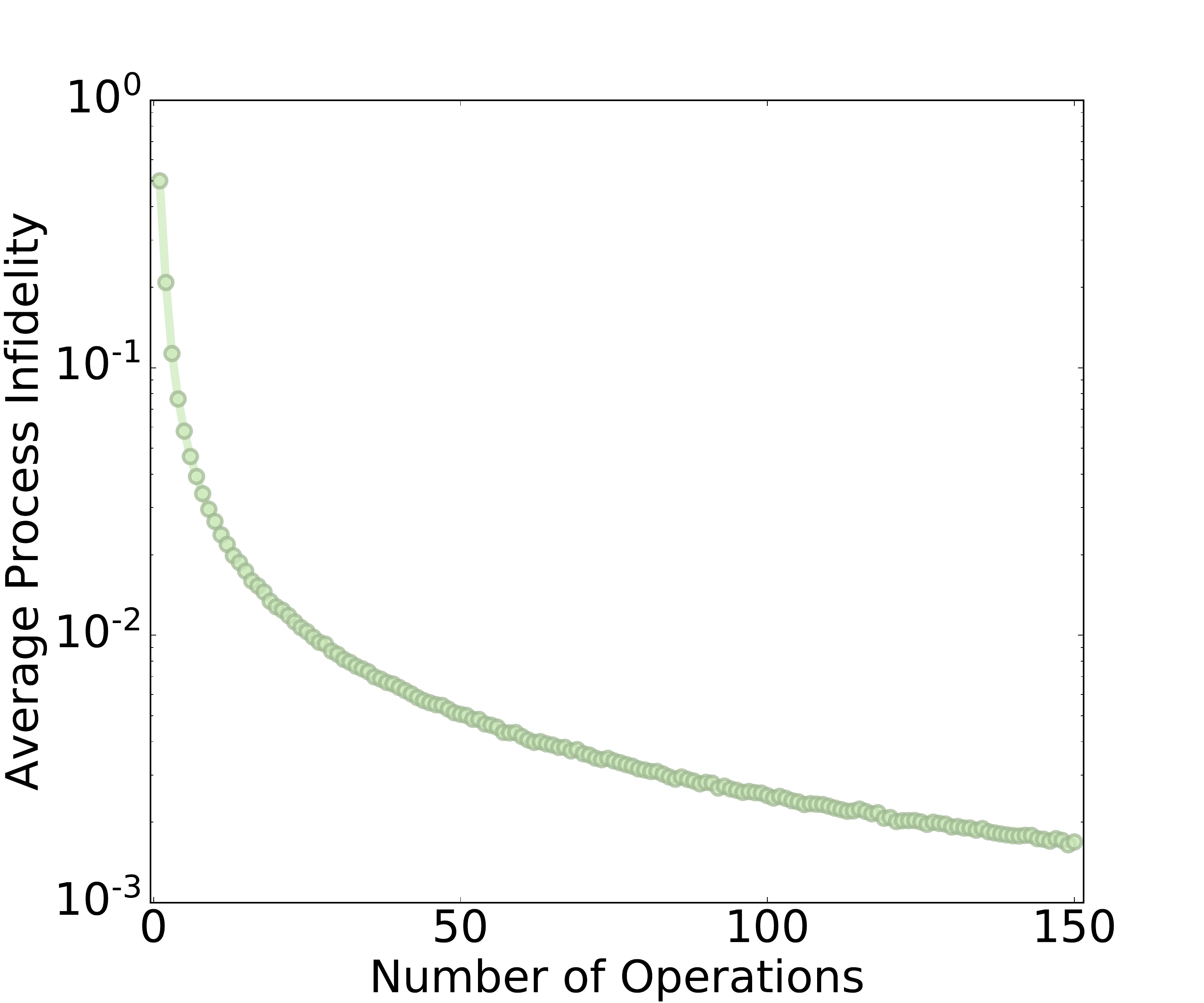}
  \caption{\textbf{Monte Carlo simulation of the BB84-channel with $p=0.5$.} A plot of the average process infidelity between the ideal process and the simulated process versus the number of applied operations used to simulate the noisy channel. The infidelity is defined as $1-F_{\text{av}}$, where $F_{\text{av}}$ is the fidelity. Hence, smaller infidelities indicate a higher degree of agreement.}
    \label{fig:bb84MCsim}
\end{figure}

As described in the main text, we implemented the noisy channels in two different ways.
In this section, we will discuss the first method, wherein we randomly apply either a Pauli-$X$, -$Y$, -$Z$ or the identity operation for one second of our data acquisition time.
The probability of each operation is given by the type of noisy channel we wish to implement (\textit{i.e.}, by one of Eqs.~\eqref{XY}--\eqref{BB84}).
The natural question is how many operations must we average over to ensure a faithful implementation of the noisy channels.

To answer this, we used Monte Carlo simulations to study the average `process fidelity' $F_{\text{av}}$ as a function of the number of applied operations.
We computed the average process fidelity (defined in Ref.~\cite{gilchrist2005distance}) by \textit{i.}\ randomly generating $10000$ single qubit states from the Haar measure, \textit{ii.}\ computing the ideal output state $\rho_\mathrm{id}$ using Eqs.~\eqref{XY}--\eqref{BB84}, \textit{iii.}\ simulating the output by applying $N$ randomly chosen operations $\rho_\mathrm{sim}$, and then \textit{iv.}\ computing the average fidelity between $\rho_\mathrm{id}$ and $\rho_\mathrm{sim}$ for all $N$ input states. Already for $N=25$, the average process fidelity is larger than $0.99$.
As an example, a plot of the average `process infidelity' $(1-F_{\text{av}}$) for the BB84-channel with $p=0.5$ is shown in Fig.~\ref{fig:bb84MCsim}.
(For the infidelity, a value of $0$ indicates a perfect implementation.)
We chose this as a representative example since the BB84-channel takes a slightly longer time to converge than all the others (this is because the BB84-process randomly applies one of the $4$ operations, while all the others only choose among $2$ operations). So, the case shown in Fig.~\ref{fig:bb84MCsim} represents the worst case among all the ones studied.
Nevertheless, even such a channel converges to the ideal noisy channel quite rapidly with $N$.
Finally, since we always implement two channels simultaneously, we apply $1000$ different operations, which is far beyond this limit.

\subsection{Single-Photon Source}
\label{subsec:Source}

A CW laser centered at 392nm emits the pump beam for a source producing single photons through a process of type-II spontaneous-parametric-down-conversion. The pump beam traverses a focusing lens with $f=\text{12.5}$cm, and then reaches, at the proper distance, a 3mm-thick beta-barium borate (BBO) crystal. Within the crystal, single photons are generated at a wavelength centred at 784nm. To compensate for the spatial and temporal walk-off of the resulting single photon pairs, they are sent each through a BBO crystal of  1.5mm thickness. They are finally filtered in polarization through a longpass filter, and a bandpass filter centered at 785nm with a full-width-half-maximum of 10nm. The photon pairs rate is 23000/s with a pump power of 85mW.

\subsection{Quantum Process Tomography}
\label{subsec:process_tomography}

Our experimental measurements consist, in general, of performing two-qubit QPT on a path and a polarization qubit.
Basically, QPT requires two steps, \textit{i.}\ preparing the system in a tomographically-complete set of states before the process, and \textit{ii.}\ measuring the system in a complete basis set after the process.
For the polarization qubit, this is relatively straightforward.
In fact, in all of three superposition methods outlined in Fig.~\ref{setup}, the photons enter the experiment in a single path.  
At this point a QWP and a HWP are inserted, which allow us to prepare any single-qubit polarization state.
After this, the path qubit is prepared by a 50/50 beamsplitter in a quantum superposition of two paths.
After the noisy channels, the paths are recombined by another 50/50 beamsplitter. On each of the output paths we place a QWP, a HWP and a polarizing beamsplitter to implement the polarization measurements.
Although they are physically different elements, we ensure that the waveplates in each output arm are always set to the same angle, and hence perform the same measurement.

Even though, in all of the communication schemes presented in the main text, the trajectory is simply initialized in an equal superposition (\textit{i.e.}, the path qubit starts out in $\ket{+}_\mathrm{T}$), one must prepare this qubit in a complete set of states in order to perform QPT. 
In order to change the input state of the path qubit between $\ket{+}_\mathrm{T}$, $\ket{R}_\mathrm{T}=\bigl(\ket{0}_\mathrm{T}-i\ket{1}_\mathrm{T}\bigr)/\sqrt{2}$, and $\ket{L}_\mathrm{T}=\bigl(\ket{0}_\mathrm{T}+i\ket{1}_\mathrm{T}\bigr)/\sqrt{2}$, we set the relative phase between the two trajectories after the first beamsplitter using a delay stage mounted on a calibrated piezo-actuator.
We can also easily prepare $\ket{0}_\mathrm{T}$ and $\ket{1}_\mathrm{T}$ by blocking either path.
Analogously, we measure the path qubit in two different ways.
To measure in $\{\ket{+}_\mathrm{T},\ket{-}_\mathrm{T}\}$, or $\{\ket{R}_\mathrm{T},\ket{L}_\mathrm{T}\}$, we suitably set the relative phase between the two paths before recombining them at the second beamsplitter.
We use the same delay stage to both set the phase of the path state, and to measure it in $\{\ket{+}_\mathrm{T},\ket{-}_\mathrm{T}\}$, or $\{\ket{R}_\mathrm{T},\ket{L}_\mathrm{T}\}$.
This can be done by adding the required phase for state preparation and subtracting the phase for state measurement. Such a phase is then converted into a path delay and sent to the piezo-actuated delay stage.
To measure in the $\{\ket{0},\ket{1}\}$ basis, we block either path before the 50/50 beamsplitter, and we then sum the counts from the two paths after the beamsplitter. 

To collect a complete set of data, we prepare the path qubit in $\{\ket{0}_\mathrm{T}, \ket{+}_\mathrm{T}, \ket{R}_\mathrm{T}, \ket{L}_\mathrm{T}\}$, and for each of these path states we prepare the polarization qubit in $\{\ket{0}_\mathrm{I}, \ket{+}_\mathrm{I}, \ket{R}_\mathrm{I}, \ket{L}_\mathrm{I}\}$, for a total of 16 input states.
We then measure each of these 16 two-qubit states by setting 9 different two-qubit basis settings: $\bigl\{\ket{0,0},\ket{0,+},\ket{0,R},\ket{+,0},\ket{+,+},\ket{+,R},\ket{R,0},\ket{R,+},$ $\ket{R,R}\bigr\}_\mathrm{I,T}$.
However, for each measurement setting we measure all four outcomes.
For example, when the measurement is set to $\ket{0,0}_\mathrm{I,T}$, we obtain the projections onto $\ket{0,0}_\mathrm{I,T}$, $\ket{0,1}_\mathrm{I,T}$, $\ket{1,0}_\mathrm{I,T}$, and $\ket{1,1}_\mathrm{I,T}$.
This yields 36 different measurement results for each of the 16 input states, providing an over-complete data set, on which we perform a least-squares QPT routine.

Equipped with this mathematical description of our experimental channel, we can compute the action of our experiment on one qubit of a maximally-entangled Bell state when the path qubit is set to $\ket{+}_\mathrm{T}$.
From this, we evaluate the coherent information (Eq.~\eqref{eq:Ic}).
Fixing the state of the path qubit in this manner results in the coherent information of the effective one-to-two-qubit channel.

We carry out this method based on full QPT for the cases of quantum-control of parallel channels and channels in series with quantum-controlled operations, but for the quantum-control of channel order we can make a simplification to lower bound the coherent information which saves significant measurement time.
For these data, we only prepare the path state $\ket{+}_\mathrm{T}$, and then measure it in the $\{ \ket{+}_\mathrm{T},\ket{-}_\mathrm{T} \}$ basis, as described above.
With these measurements, we lower bound the coherent information in our channels as follows.
We first reconstruct two single-qubit $\chi$-matrices for the target systems, $\chi^{\ket{+}}$ and $\chi^{\ket{-}}$, using single-qubit process tomography on the polarization qubit. In particular,
$\chi^{\ket{+}}$ is the single-qubit effective process that the information qubit experiences when the trajectory measurement results is $\ket{+}_\mathrm{T}$, whereas $\chi^{\ket{-}}$ is the effective process when the trajectory measurement outcome is $\ket{-}_\mathrm{T}$  

Next, we compute the action of the one-qubit $\chi$-matrix on a maximally-entangled Bell state, to evaluate the two values of the coherent information $\mathcal{I}_c^{\ket{+}}$ and $\mathcal{I}_c^{\ket{-}}$ in Eq.~\eqref{eq:Ic}. 
Afterwards, we simply calculate their average, with each term weighted by their respective post-selection probabilities $p_{\ket{+}}$ and $p_{\ket{-}}$:
\begin{equation}
\mathcal{I}_c^\mathrm{LB} = p_{\ket{+}} \mathcal{I}_c^{\ket{+}} + p_{\ket{-}} \mathcal{I}_c^{\ket{-}}.
\end{equation}
In general, $\mathcal{I}_c^\mathrm{LB}$ sets a lower bound on $\mathcal{I}_c$ because of the data processing inequality for coherent information \cite{Schumacher1996, nielsen2000quantum}.
Furthermore, in absence of additional errors, it can be shown that $\mathcal{I}_c^\mathrm{LB}=\mathcal{I}_c$ in the case of the quantum switch.

\subsection{Fixing the Quantum-Controlled Operations Independently of the Noise}
\label{subsec:Fixing_Operations}

\begin{figure}
  \centering%
      \includegraphics[width=\columnwidth]{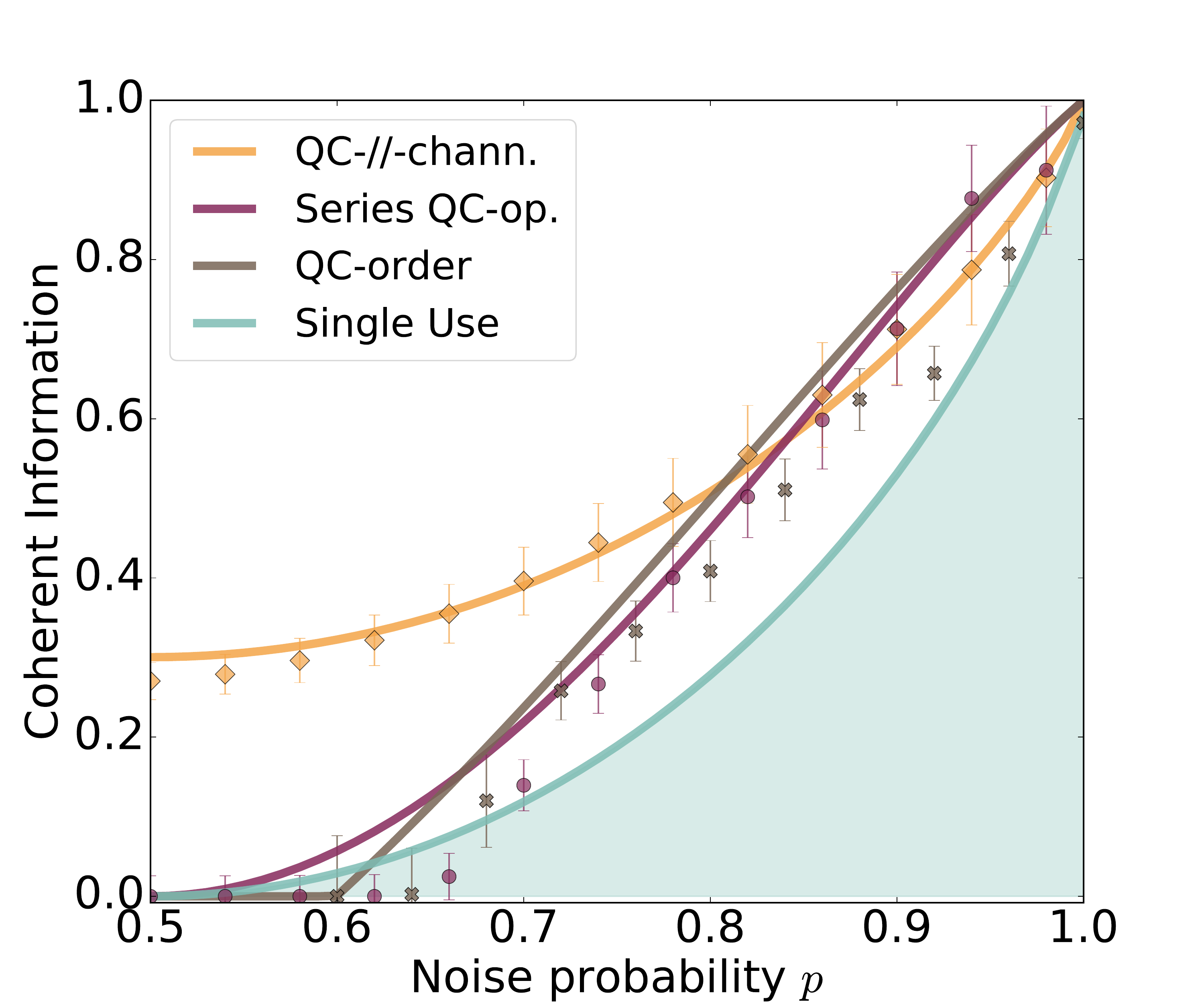}
  \caption{\textbf{Experimental BF- and PF-noise data for sub-optimal quantum-controlled operations.} The trend of the scheme featuring the channels in series with quantum-controlled operations (Series QC-op.) performs worse than the quantum-control of channel order (QC-order) for all $p \geq 0.67$, but better than the quantum-control of parallel channels (QC-//-chann.) for $p \geq 0.84$. 
  The experimental data for the sub-optimal choice of Series QC-op. is in good agreement with the expected trend.}
    \label{fig:BitPhasechan_withId}
\end{figure}

In some cases, for instance in a rapidly-varying noise environment, it may be impossible to estimate the type of noise and adapt the quantum-controlled operations accordingly. In these situations, one would need to fix such operations independently of the noise. For our noise varieties, the operations $U_2$ and $U_3$ were set to $\mathcal{I}$ in the cases of the \textit{XY}- and BB84-channels, and to $H$ for the BF-and-PF case, whereas $U_1 = Y$ in all three cases. If we were to keep the same quantum-controlled operations in the BF-and-PF case as in the \textit{XY} and BB84 cases, the efficiency of the scheme would be reduced, and the channel activation due to the channels in series with quantum-controlled operations would result comparable to that of the two other schemes (\textit{i.e.}, the quantum-control of parallel channels, and the quantum-control of channel order). The theoretical trend and the experimental data points corresponding to this case are shown on Fig.~\ref{fig:BitPhasechan_withId}. Colors and data points shapes are the same as in Figs.~\ref{fig:XYchan}--\ref{fig:BB84}.

\bibliography{references}

\end{document}